\documentclass{IEEEtaes}

\usepackage{color,array,amsthm}
\usepackage{graphicx}
\usepackage{xcolor}
\usepackage[colorlinks,urlcolor=blue,linkcolor=blue,citecolor=blue]{hyperref}

\usepackage{amssymb}
\usepackage{amsmath}
\usepackage{amsfonts}
\usepackage{cite}
\usepackage{microtype}
\usepackage[version=4]{mhchem}
\usepackage{siunitx}
\usepackage{bm}
\usepackage{algorithm}
\usepackage{algpseudocode}
\usepackage{longtable,tabularx}
\usepackage{multirow}
\usepackage{subfigure}
\usepackage{caption}
\usepackage{placeins}
\usepackage{float}
\usepackage{nomencl}
\usepackage{booktabs}

\def\bf{\textbf}
\def\be{\begin{equation}}
\def\ee{\end{equation}}

\jvol{XX}
\jnum{XX}
\jmonth{XXXXX}
\paper{1234567}
\pubyear{2023}
\doiinfo{TAES.2023.Doi Number}

\newtheoremstyle{mystyle}{12pt}{12pt}{\scshape}{0cm}{}{}{0.5 em}{} 
\theoremstyle{mystyle}
\newtheorem{theorem}{\scshape Theorem}
\newtheorem{lemma}{\scshape Lemma}
\newtheorem{remark}{\scshape Remark}
\newtheorem{assumption}{\scshape Assumption}
\newtheorem{condition}{\scshape Condition}
\newtheorem{proposition}{\scshape Proposition}

\newcommand{\YL}[1]{{#1}}
\newcommand{\YLRV}[1]{{#1}}

\newcommand{\YLnew}[1]{{#1}}

\newcommand{\alphak}{{\bm{\alpha}[k]}}
\newcommand{\alphakfirst}{{\bm{\alpha}[k-1]}}
\newcommand{\Pk}{{\bm{P}[k]}}
\newcommand{\Pkfirst}{{\bm{P}[k-1]}}
\newcommand{\Phik}{{\bm{\Phi}[k]}}
\newcommand{\Phikfirst}{{\bm{\Phi}[k-1]}}
\newcommand{\rhok}{{\bm{\rho}[k]}}
\newcommand{\rhokfirst}{{\bm{\rho}[k-1]}}
\newcommand{\Lk}{{\bm{L}[k]}}

\newcommand{\yi}{{\bm{y}[i]}}
\newcommand{\tildexi}{{\tilde{\bm{x}}[i]}}
\newcommand{\tildexj}{{\tilde{\bm{x}}[j]}}
\newcommand{\tildexk}{{\tilde{\bm{x}}[k]}}
\newcommand{\tildext}{{\tilde{\bm{x}}[\tau]}}
\newcommand{\tildexfirst}{{\tilde{\bm{x}}[1]}}
\newcommand{\tildexN}{{\tilde{\bm{x}}[N]}}
\newcommand{\yjfirst}{{\bm{y}_j[1]}}
\newcommand{\yjN}{{\bm{y}_j[N]}}
\newcommand{\xk}{{\bm{x}[k]}}
\newcommand{\uk}{{\bm{u}[k]}}
\newcommand{\yk}{{\bm{y}[k]}}
\newcommand{\eek}{{\bm{\epsilon}[k]}}

\newcommand{\TR}[1]{{#1}}

\newcommand{\LY}[1]{{#1}}

\newcommand{\RV}[1]{{#1}}

\begin{document}

\title{Attitude Takeover Control for Noncooperative Space Targets Based on Gaussian Processes with Online Model Learning \vspace{-12mm}}

\author{YUHAN LIU}
% \member{Member, IEEE}
\affil{Eindhoven University of Technology, Eindhoven, The Netherlands} 

\author{PENGYU WANG}
% \member{Member, IEEE}
\affil{Korea Advanced Institute of Science and Technology, Daejeon, Korea} 

\author{CHANG-HUN LEE}
% \member{Member, IEEE}
\affil{Korea Advanced Institute of Science and Technology, Daejeon, Korea} 

\author{ROLAND T\'OTH}
% \member{Senior Member, IEEE}
\affil{Eindhoven University of Technology, Eindhoven, The Netherlands\\
Institute for Computer Science and Control, Budapest, Hungary}

 % \receiveddate{Manuscript received XXXXX 00, 0000; revised XXXXX 00, 0000; accepted XXXXX 00, 0000.\\
% This paragraph of the first footnote will contain the date on which you submitted your paper for review, which is populated by IEEE. It is IEEE style to display support information, including sponsor and financial support acknowledgment, here and not in an acknowledgment section at the end of the article. For example, ``This work was supported in part by the U.S. Department of Commerce under Grant BS123456.'' }
%% \accepteddate{XXXXX XX XXXX}
%% \publisheddate{XXXXX XX XXXX}

\receiveddate{This research has received funding from the European Union’s
EDF programme under grant agreement No 101103386 and has been supported by the European Union within the framework of the National Laboratory for Autonomous Systems (RRF-2.3.1-21-2022-00002).}

\corresp{\\ {\itshape(Corresponding author: Pengyu Wang)}}

\authoraddress{Yuhan Liu is with the Control Systems Group, Department of Electrical Engineering, Eindhoven University of Technology, Eindhoven, The Netherlands. (Email: \href{y.liu11@tue.nl}{y.liu11@tue.nl}); 
Pengyu Wang and Chang-hun Lee are with the Flight Dynamics and Control Laboratory, Department of Aerospace Engineering, Korea Advanced Institute of Science and Technology, 
Daejeon, Korea. (Email: \href{wangpy@kaist.ac.kr}{wangpy@kaist.ac.kr}, \href{lckdgns@kaist.ac.kr}{lckdgns@kaist.ac.kr}); 
Roland T\'oth is with the Control Systems Group, Department of Electrical Engineering, Eindhoven University of Technology, Eindhoven, The Netherlands and the Systems and Control Laboratory, Institute for Computer Science and Control, Budapest, Hungary. (Email: \href{r.toth@tue.nl}{r.toth@tue.nl}).}

% \editor{Mentions of supplemental materials and animal/human rights statements can be included here.}

\markboth{YUHAN LIU ET AL.}{ATTITUDE TAKEOVER CONTROL FOR NONCOOPERATIVE TARGETS}
\maketitle \vspace{-2mm}

\begin{abstract}One major challenge for autonomous attitude takeover control for on-orbit servicing of spacecraft is that an accurate dynamic motion model of the combined vehicles is highly nonlinear, complex and often costly to identify online, which makes traditional model-based control impractical for this task. To address this issue, a recursive online sparse Gaussian Process (GP)-based learning strategy for attitude takeover control of noncooperative targets with maneuverability is proposed, where the unknown dynamics are online compensated based on the learnt GP model in a semi-feedforward manner. \YL{The method enables the continuous use of on-orbit data to successively improve the learnt model during online operation and has reduced computational load compared to standard GP regression. Next to the GP-based feedforward, a feedback controller is proposed that varies its gains based on the predicted model confidence\TR{,} ensuring robustness of the overall scheme.}
Moreover, rigorous theoretical proofs of Lyapunov stability and boundedness guarantees of the proposed method-driven closed-loop system are provided in the probabilistic sense. \YL{A simulation study based on a high-fidelity simulator is used to show the effectiveness of the proposed strategy and demonstrate its high performance.} 
\end{abstract}
\begin{IEEEkeywords}Attitude takeover control, Gaussian process, machine learning, noncooperative space target, on-orbit servicing. \vspace{-6mm}
\end{IEEEkeywords}

\section{INTRODUCTION}
\label{sec:intro}
I{\scshape n} recent years, there has been a rapid development in \YL{\emph{on-orbit servicing} (OOS) }applications such as on-orbit refueling, on-orbit maintenance, on-orbit assembly, orbit transfer, and active space debris removal \cite{flores2014review}. An entire OOS mission %can be 
\TR{consists of} %segmented 
%into 
the following four distinct phases: long-range guidance, final approaching, on-orbit capture, and post-capture. In the post-capture phase, a combined spacecraft is formed through the connection of the servicing spacecraft (referred to as the ``servicer") and the target using robot manipulators or tethers. \RV{%Meanwhile, 
\TR{The} servicer \TR{is supposed to} %will 
``\emph{take over}" the attitude control of the target, 
%wherein 
\TR{such that} %and 
the control torque for the combined spacecraft is completely provided by the servicer. }\TR{This} autonomous attitude takeover control for the combined spacecraft plays a key role in the subsequent tasks (such as refueling and debris removal) and has become an %urgent 
\TR{important component} %concern 
to ensure the success of OOS missions.

\TR{C}apture and post-capture control for cooperative targets have significantly matured and have been applied in \TR{some} executed  OOS missions \cite{pinard2007accurate,liu2018key}.
%for instance, Orbit Express \cite{friend2008orbital} in the US, ATV \cite{pinard2007accurate} in Europe, ETS-VII \cite{oda2000experiences} in Japan, and TianGong-2 \cite{liu2018key} in China. 
However, for other OOS missions, such as on-orbit maintenance and debris removal, the targets are usually noncooperative and the mission needs to be conducted under the following specs: 1)  sufficient knowledge of the structure of the target, mass properties, and state of motion \TR{is not \emph{a priori} available}; 2) \TR{n}o communication link can be established to send messages between the servicer and target; 3) \TR{n}o pre-designed capture interface \TR{is present} on the target. With the increasing diversity of OOS missions, the targets can also display partial failure characteristics, i.e., still have weak attitude controllability despite the failure of actuators. Hence, for upcoming OOS missions, the attitude control for the post-capture combined spacecraft %is 
\TR{is likely be} more challenging. % than for a single spacecraft. 
First, noncooperative characteristics of the target in terms of 2) necessitate robustness and disturbance rejection capabilities of the servicer. \RV{Furthermore, in view of 1) and 3), no accurate model can be assumed to be available for the dynamics of the combined spacecraft. 
Hence, traditional model-based control %design 
methods \TR{can} %which 
suffer from \TR{the unknown} uncertainties and unmodeled dynamics \TR{that can be encountered during such missions}, \TR{and can result in significant loss of performance or even stability}.} %are inapplicable for the combined spacecraft due to the performance deterioration.}
% which results in the inapplicability of the traditional model-based control design methods. 
Additionally, it is not feasible to effectively identify the mass properties of the combined spacecraft in real-time due to the external unmeasured input caused by the attitude maneuverability of the target.

% claim the difficulties for attitude takeover control for noncooperative targets (theoretical aspect)
In scientific literature, already promising studies have been obtained for post-capture attitude takeover control. The pioneering work in this field \TR{has} mainly concentrated on designing model-based controllers on the basis of an accurately identified model. For this purpose, Bergmann \emph{et al.} \TR{in} \cite{bergmann1987mass} \TR{have} proposed an online inertia identification algorithm, where the mass properties of a rigid spacecraft \TR{are} estimated by the analytic solution of the \TR{motion} equation under free rotation. \RV{In \cite{murotsu1994parameter}, Murotsu \emph{et al.} \TR{have} developed a parameter identification method based on the conservation of momentum. %theorem 
%with the assistance of robot arms. 
Ma \emph{et al.} \TR{in} \cite{ma2008orbit} \TR{have} further extended this work to scenarios %with 
\TR{under} unknown \YLRV{spacecraft \TR{systems}} 
%flying bases
by a two-step identification method. Christidi-Loumpasefski  \emph{et al.} \TR{in} \cite{christidi2023parameter,christidi2023system} \TR{have} proposed momentum-conservation-based methods to fully identify the parameters of free-flying system dynamics with unmeasurable sloshing states. %, which can be applied to multi-arm systems. 
}
%The results were then enhanced to be concurrently utilized with a transpose Jacobian controller \cite{christidi2020concurrent}. }
 %A quasi model-free attitude control strategy with an online database for combined spacecraft was developed in \cite{she2018quasi}, where the corresponding controllers were designed for different target models identified by the virtual coordinate method.
 In recent years, visual CCD cameras \cite{meng2019identification} and deep learning \cite{chu2020least} have also been proposed for \TR{estimation} %the identification 
 of inertia parameters of the combined spacecraft. However, %performance of identification-based approaches \TR{in general} is \TR{significantly} affected by the presence of measurement and space noise. Furthermore,  
 these methods are not applicable to %the task 
 scenario\TR{s} where the target still has attitude maneuverability.  

% Besides data-driven modeling and estimation, 

\RV{In contrast to parameter identification and model-based control approaches,
an effective alternative approach to deal with the unknown dynamics of the combined spacecraft is adaptive control.} %, which offers an attractive alternative for attitude control for the combined spacecraft. 
In \cite{huang2015adaptive}, a backstepping-based robust adaptive controller \TR{has been} %was
proposed for attitude stabilization of a tumbling tethered combined spacecraft. By involving an improved adaptive sliding mode controller to reduce the total angular momentum of the system, Zhang \emph{et al.} \TR{in}  \cite{zhang2017coordinated} \TR{have} presented a coordinated control approach for the combined spacecraft without precise inertia information. 
Kang \emph{et al.} \TR{in} \cite{guohua2020adaptive} \TR{have} considered the situation of non-cooperative body attachment, and designed an adaptive control strategy for \TR{rapid} stabilization with %fast speed and 
high precision in different scenarios. 
%In \cite{wang2017adaptive}, an adaptive compensation module was introduced into the feedback control law based on an adaptive fault tolerance approach.
A hybrid %non-fragile 
controller \TR{has been} %was 
derived in \cite{liu2021nonfragile} for %the 
\TR{a} flexible combined spacecraft in the presence of model uncertainties, input constraints, external disturbances, and actuator faults. %, which achieved vibration suppression while stabilizing the attitude of the combined spacecraft.
%A concurrent learning technique was utilized in \cite{zhao2021concurrent} to develop an adaptive finite-time control law for desired trajectory tracking. 
Moreover, to further improve the adaptability of the controller, \emph{neural network} (NN) and fuzzy logic-based control approaches have attracted great attention recently (see,
e.g., \cite{LEEGHIM2009778,huang2016impact,ning2021event}). 
However, it is usually essential for adaptive control approaches to make assumptions on the existence of upper bounds on the model uncertainty, external disturbances, etc. to ensure the stability of the closed-loop system, which \TR{can be} conservative for an attitude takeover task. On the other hand, parametric modeling methods such as NN and fuzzy logic inherently have the disadvantages of {being complex and fragile in terms of their generalization capability beyond the training region, and can be sensitive in terms of the selected network structure, activation functions, hyperparameter tuning, initialization, and signal normalization.}

%still dependent on model information for the selection of the activation functions. For instance, in the NN-based approach, the center position and width of the activation function need to be appropriately selected according to the model information, and the system state needs to be kept near the center of the activation function to ensure that the unknown function is sufficiently excited and accurately captured.

Note that the above references involve system models to facilitate the controller design, i.e., model-based control. In contrast, there are some results of ``model-free" control for the combined spacecraft. In \cite{wei2018adaptive}, a model-free prescribed performance adaptive attitude controller \TR{has been} proposed for the flexible combined spacecraft dynamics. In \cite{luo2018low}, a low-complexity model-free control strategy \TR{has been} presented {based on the} %the framework of the 
prescribed performance technique. 
%A concept of ``passable performance bound" was %further 
%{included in this approach}
%introduced into this framework 
%in \cite{hu2020model} to {improve the tuning of the}
%well-tune the 
%control parameters.
In \cite{fan2021inertia}, an inertia-free control approach \TR{has been} derived for {the} combined spacecraft which facilitates  %ted the 
``appointed-time'' stability of the system. \YL{However, it should be mentioned that these model-free control strategies are {implicitly dependent on the assumed inertia of the system through their hyperparameter choices}. %  still lack generalization {cap}ability {[I do not get the meaning of the next part. Should it be removed?]} though no inertia information involves in the controllers explicitly. 
{Hence, the}
%Furthermore,  many 
parameters in these model-free control laws need to be tuned based on model information and practical experience, which makes the ``model freeness" of these methods questionable.}
% First, they generally require lots of assumptions such as model form (i.e., affine, Euler-Lagrange) and boundedness of uncertainties to avoid model-related terms involved in the stability proof. Second, 

% In view of all the developed approaches mentioned above, \YL{a variety of control laws were developed by directly utilizing the system input/output data}, namely the data-driven control. As a typical data-driven control method, \YL{PID control with gain tuning} has been widely applied to the attitude control of combined spacecraft \cite{stolfi2017combined,li2018dynamics,chu2018hybrid}. Besides, other direct data-driven control methods such as \emph{model-free adaptive control} (MFAC), \emph{virtual reference feedback control} (VRFC), \emph{iterative feedback rectification control} (IFRC), etc. have been utilized as control strategies for combined spacecraft in \cite{xie2018data,gao2019forecasting,han2019data}. Although direct data-driven control methods have been extensively studied in recent years, many of these results only exploit the "transient" feature of the algorithms, but ignore to take the advantage of "data" in controller design, that is, learning like a human being: extracting valid "knowledge" and "experience" from the system data to improve the intelligence of the controller.

\YL{In view of all the aforementioned developed approaches and the identified challenges, the current shortcomings in the area of combined
spacecraft attitude takeover control are summarized as: }

\YL{\TR{(\emph{i}}) General identification-based control approaches are not applicable to the task scenario where the target still has attitude maneuverability (low frequent unknown excitation).}

\YL{\TR{(\emph{ii}}) {Performance of adaptive control methods is highly sensitive to hyperparameter choices.} % highly depends on the %ler design. As their performance is highly sensitive to the obtained choices. 
Inadequate selection may lead to degradation of the overall control performance and even loss of stability.}

To address these issues, the emerging approaches of machine learning-based control offer better capabilities for capturing unmodeled dynamics and achieving superior performance over the existing methods. \YL{As one of the promising tools, GP \cite{rasmussen2003gaussian} has been increasingly successful in the field of nonparametric modeling. It is a flexible function estimator that also provides a characterization of the uncertainty of the estimate in a computationally efficient manner compared %for example 
to NNs and fuzzy logics.}
Powerful results have been achieved for GP-based learning control for robot arms \cite{beckers2019stable}, quadrotors \cite{liu2021learning}, race cars \cite{kabzan2019learning}, etc. 
\YL{However, there are still some technical barriers to the design of GP-based learning control for the attitude takeover problem:
\YLRV{(a)} The standard GP is not suitable for large training data sets, which will lead to high computational load for the onboard computer. 
\YLRV{(b)} Most GP-based learning control methods do not perform online updating which is needed to address the time-varying disturbances during the system operation.
}
\RV{To solve this issue, various online GP approaches have been developed to update the GP model during control operation.  A sparse online GP (SOGP) method \TR{has been} first proposed in \cite{csato2002sparse}, which efficiently approximates the Kullback-Leibler divergence, i.e., the distance, between the current GP model and the new data pair and updates the \TR{GP estimate} %budgeted basis vector 
based on this distance. This work \TR{has been} further extended and applied in model reference adaptive control \cite{grande2014experimental,chowdhary2014bayesian}, incremental backstepping control \cite{ignatyev2023sparse}, etc. Similarly, another online GP \TR{approach} %widely used in the control field 
is evolving GP \cite{petelin2011control,kocijan2016modelling}, which updates the training data set online using \TR{various} %some 
type\TR{s} of information \TR{criteria},  %on, 
and has been utilized in model predictive control \cite{maiworm2021online}. However, both of the above methods still involve  \TR{a} ``dictionary" update and re-computation of the Gram matrix at each time step, \TR{which is a computationally costly operation}.}

To overcome the aforementioned challenges, this paper proposes an innovative GP-based online learning strategy for post-capture attitude takeover control with unknown dynamics and attitude maneuverability of the target, as an extension of our previous work presented in \cite{ma2021learning}. The main contributions of this paper are as follows: %\YLcom{Please check item 2) here}
\begin{itemize} 
\item[1)] {A novel GP-based learning control strategy for attitude takeover that is applicable even under attitude maneuverability of the target. }
%\YL{A novel recursive online sparse GP algorithm is presented to continuously learn the unknown time-varying uncertainties during control operation. The crucial feature is that there is no need to re-optimize the hyperparameters or update the "dictionary" at every time moment, which facilitates real-time updating of the GP model with online streaming data while the computational load is kept at a low level.}

% \vspace{0.1 cm}\item[2)]
% \YLnew{As a data-driven approach, the proposed GP-based online learning control strategy avoids costly modeling and identification of the combined spacecraft.} 
\item[2)] {A novel recursive online sparse form of the GP estimator that facilitates efficient continuous learning of unknown time-varying uncertainties during operation. A crucial advantage of the approach is that no online re-tuning of the hyperparameters, nor update of the data-dictionary is required at every time-moment, which ensures low online computational cost.}

%{\YLnew{A novel GP-based online learning control strategy is proposed for the attitude takeover task where the target remains the attitude maneuverability. Additionally, the rigorous Lyapunov-based stability analysis in the sense of probability is derived, where the system state error is guaranteed to be ultimately bounded within a small set containing the origin with high probability.}}

% it is enough to have an initial "rough" model of the system space before the mission execution, which initial model is updated successively to ensure gradual improvements in the accuracy of the corresponding control algorithm.

%\vspace{0.1 cm}
\item[3)] {Proven stability guarantee of the closed-loop operation with the proposed method, ensuring that the attitude orientation error remains ultimately bounded around the origin with high probability.}
%\YL{Distinct from  most existing works, the selection for controller parameters is significantly simplified without  "trial and error" step or subjective practical experience. This feature enhances the generality and flexibility of the whole control strategy.}

%\vspace{0.1 cm}
\item[4)] {Verification of the proposed method in a high-fidelity simulation study.}
%\YL{The effectiveness of our proposed control strategy is verified by numerical simulations on a high-fidelity simulation platform.}
\end{itemize}

\YL{The remainder of this paper is organized as follows. Section II covers the problem formulation and control objectives. 
The proposed recursive online sparse GP regression algorithm is detailed in Section III.
The GP-based adaptive learning control procedure and its rigorous stability analysis are presented in Section IV. Numerical simulation results are provided in Section V, followed by the conclusions in Section VI.}
\vspace{2mm}

\YLnew{\emph{Notation}: %Throughout this paper, 
\RV{$\mathbb{R}$, $\mathbb{Z}$, {$\mathbb{N}$}, and ${\mathbb{Q}}^3$ denote the sets of real numbers, integers, {nonnegative integers}, and unit quaternions, {$\mathbb{R}_0^+$ corresponds to nonnegative real numbers, while $\mathbb{S}^{n\times n}$ is the set of real symmetric matrices of dimension $n \times n$}.
The 2-norm of a vector or a matrix is {denoted} as $\|\cdot\|$, while, {for a given Hilbert space $\mathcal {H}$,} the {corresponding} %\emph{reproducing kernel Hilbert Space} 
 %(RKHS) 
 norm is denoted by $\|\cdot\|_{\mathcal{H}}$.
%We denote by  
$\lambda_{\rm{min}}(\cdot)$ and $\lambda_{\rm{max}}(\cdot)$ {are}  the minimum and maximum eigenvalues of a matrix, respectively. %Let 
$\mathrm{vec}(\bm{x}_1,...,\bm{x}_n)=[\bm{x}_1^{\top} \ \cdots \ \bm{x}_n^{\top}]^{\top}$ denotes the column-wise composition of vectors. Additionally, $\bm{I}_{n}$ {is} %denotes 
a $n\times n$ identity matrix and {the projection} $\TR{\centerdot}^{\times}:\mathbb{R}^{3}\to\mathbb{R}^{3\times 3}$  {gives} %ndicates 
a skew-symmetric matrix, % for any given vector $\bm{a} = [a_1~ a_2~ a_3]^{\top}$.
{ensuring that $\bm{a}^{\times}\bm{b} = \bm{a}\times \bm{b}$ for all $\bm{a},\bm{b}\in \mathbb{R}^3$ where $\times$ corresponds to the cross-product operator. $\mathbb{I}_\mathrm{I}^{j}=\{s\in\mathbb{Z} \mid i \leq s \leq j\}$ denotes an index set.}
}}

\section{Problem Formulation}
\label{sec:2}
\subsection{Uncertain System Model}
\label{sec:sub:model}
\begin{figure*}[htb]
  \centering\includegraphics[width= 5 in]{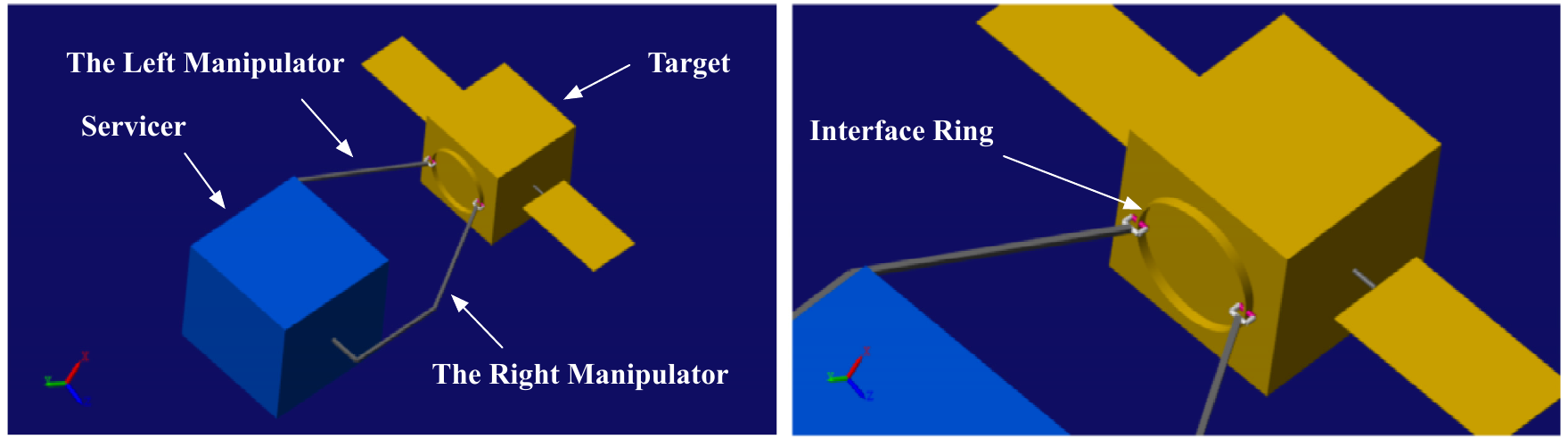} \vspace{2mm}
  \caption{A snapshot of the combined spacecraft} \vspace{-3mm}
  \label{fig:1}
\end{figure*}

{After successful docking to the target, the} combined spacecraft, as considered in this paper, contains three parts: the servicer, the target, and the manipulators. As shown in Fig.~\ref{fig:1}, the servicer can capture the interface ring on the target by using its two manipulators. \RV{\TR{To describe the motion dynamics of the combined spacecraft, first %we will 
assume that it can be described as a single rigid body, which also means that the manipulator arms do not introduce additional dynamics.}} 
%We first derive an ideal baseline case where the combined spacecraft can be well-described as a single rigid body, where %unit 
%quaternion{s} {are used} %is adopted 
%to represent global rotations {to avoid}  %without 
%singularity. It is worth mentioning that the manipulator dynamics is not considered in this simplified case.}
This \TR{simplified} motion %differential equation describing the 
dynamics %is 
{of the combined spacecraft are} given by \cite{wie2008space}
\YLnew{
\be
\label{1}
\dot{\bm{q}}=\frac{1}{2}(q_{0}\bm{I}_{3}+\bm{q}^{\times})\bm{\omega},
\quad \dot{q_{0}}=-\frac{1}{2}\bm{q}^\top\bm{\omega},
\ee
where $\bm{Q}\!=\![q_{0}~\bm{q}^\top]^\top \in{\mathbb{Q}}^{3}$ denotes the unit quaternion describing the attitude of the spacecraft \TR{in terms of} %{, which corresponds to the 
rotation of the body frame $\mathcal{F}_\mathrm{B}$ w.r.t. the inertial frame $\mathcal{F}_\mathrm{I}$, \TR{while} $\bm{\omega}\in\mathbb{R}^{3}$ is the angular velocity of the spacecraft expressed in $\mathcal{F}_\mathrm{B}$.} %w.r.t. $\mathcal{F}_\mathrm{I}$.}

\YLnew{In this paper, \TR{the problem of} attitude orientation \TR{control} %problem 
is considered, \TR{i.e.,} %that is, 
the combined spacecraft is \TR{required} %enforced 
to \TR{realize} %rotate 
%its attitude to %a new
\TR{a commanded} desired orientation. Let $\bm{Q}_\mathrm{d}=[q_{\mathrm{d}0}~\bm{q}_\mathrm{d}^\top]^\top\in{\mathbb{Q}}^3$ 
denote the {desired} attitude{, corresponding to a desired body frame $\mathcal{F}_\mathrm{D}$, and 
$\bm{\omega}_\mathrm{d}\in\mathbb{R}^3$ the desired angular velocity,} 
%and angular velocity of the desired body frame $\mathcal{F}_\mathrm{D}$ w.r.t. $\mathcal{F}_I$, 
respectively.
 Thus, the attitude error $\bm{Q}_\mathrm{e}=[q_{\mathrm{e}0}~\bm{q}_\mathrm{e}^\top]^\top$ of $\mathcal{F}_\mathrm{B}$ w.r.t. $\mathcal{F}_\mathrm{D}$ can be calculated by $\bm{Q}_\mathrm{e}=\bm{Q}_\mathrm{d}^{*}\otimes\bm{Q}$, where $\bm{Q}_\mathrm{d}^*=[q_{\mathrm{d}0}~-\!\bm{q}_\mathrm{d}^\top]^\top$ denotes the conjugate of $\bm{Q}_\mathrm{d}$, and the symbol ``$\otimes$" refers to the %quaternion 
 product {operator for} any two quaternions $\bm{Q}_i =[q_{i0}~\bm{q}_i^\top]^\top$ and $\bm{Q}_j =[q_{j0}~\bm{q}_j^\top]^\top$: 
\begin{equation*}
\bm{Q}_i\otimes\bm{Q}_j = \left[\begin{array}{c}
q_{i0} q_{j0}-\bm{q}_i^{\top} \bm{q}_j \\
q_{i0} \bm{q}_j+q_{j0} \bm{q}_i+\bm{q}_i^{\times}\bm{q}_{j}
\end{array}\right].
\end{equation*}}

\YLnew{{The}  attitude error kinematics can be derived as:
\be
\label{2}
\dot{\bm{q}}_\mathrm{e}=\frac{1}{2}(q_{\mathrm{e}0}\bm{I}_{3}+\bm{q}_\mathrm{e}^{\times})\bm{\omega}_\mathrm{e},
\quad \dot{q}_{\mathrm{e}0}=-\frac{1}{2}\bm{q}_\mathrm{e}^\top\bm{\omega}_\mathrm{e}
\ee
where $\bm{\omega}_\mathrm{e}=\bm{\omega}-\bm{C}(\bm{Q}_\mathrm{e})\bm{\omega}_\mathrm{d}$, and the rotation matrix is %computed by
$\bm{C}(\bm{Q}_\mathrm{e})=\bm{I}_3-2q_{\mathrm{e}0}\bm{q}_\mathrm{e}^{\times}+2\bm{q}_\mathrm{e}^{\times}\bm{q}_\mathrm{e}^{\times}$. Because {the problem of set point control of} the attitude %orientation problem 
is studied in this paper, the desired angular velocity is set as $\bm{\omega}_\mathrm{d}= \mathbf{0}$, {giving} %that is, 
$\bm{\omega}_\mathrm{e}=\bm{\omega}$. Hence{,} the attitude error kinematics \eqref{2} can be written as:
\be
\dot{\bm{q}}_\mathrm{e}=\frac{1}{2}(q_{\mathrm{e}0}\bm{I}_{3}+\bm{q}_\mathrm{e}^{\times})\bm{\omega},
\quad \dot{q}_{\mathrm{e}0}=-\frac{1}{2}\bm{q}_\mathrm{e}^\top\bm{\omega}.
\ee}

\YLnew{Furthermore, the attitude dynamics {are} %can be 
given by:
\be
\bm{J}_\mathrm{c} \dot{\bm{\omega}}=-\bm{\omega}^{\times} \bm{J}_\mathrm{c} \bm{\omega}+\bm{u}+\bm{\tau}_\mathrm{d},
\ee
where $\bm{J}_\mathrm{c}\in\mathbb{R}^{3\times3}$ is the positive-definite symmetric inertia matrix, $\bm{u}\in\mathbb{R}^{3}$ is the control torque{, while} %expressed in $\mathcal{F}_\mathrm{B}$, 
\LY{$\bm{\tau}_\mathrm{d}\in\mathbb{R}^{3}$ is \TR{an} %the 
external %disturbance 
torque expressed in $\mathcal{F}_\mathrm{B}$ \TR{representing} time-varying disturbances. \TR{Note that $\bm{\tau}_\mathrm{d}$ can be seen as a collection of additional unconsidered dynamics such as manipulator arm dynamics, rotating solar panels, solar pressure, gravity gradient, etc.}}
Obviously, if $\bm{\tau}_\mathrm{d}=\bm{0}$ and $\bm{J}_\mathrm{c}$ is accurately known or identified, global asymptotic stability can be easily guaranteed {by model-based control}, %\cite{wen1991attitude}, 
\TR{ensuring}
%such 
that the closed-loop trajectory $(\bm{Q}_\mathrm{e},\bm{\omega})$ %will 
converge\TR{s} to the stable equilibrium $(\bm{1},\bm{0})$.}

%Nevertheless, 
\TR{However,}
the true value of $\bm{J}_\mathrm{c}$ is hard to determine accurately by online parameter identification when the servicer is executing an OOS mission, particularly \TR{under} %when accounting for the 
maneuvering capability of the target. Similarly, $\bm{\tau}_\mathrm{d}$ can be {partly} reconstructed by space environment models, but some residual uncertainties always remain. %Therefore, our next step is to restructure the post-capture combined spacecraft model.

%Defining 
{To be able to express some known baseline dynamics w.r.t. the real dynamics of the spacecraft, introduce}
$\bm{J}_\mathrm{c} = {\bm{J}}_{\mathrm{c}0}+\tilde{\bm{J}_\mathrm{c}}$, where ${\bm{J}}_{\mathrm{c}0}\in\mathbb{R}^{3\times3}$ represents the nonsingular symmetric nominal inertia matrix of the combined spacecraft, and $\tilde{\bm{J}_\mathrm{c}}\in\mathbb{R}^{3\times3}$ denotes the inertia deviation resulting from the capture and non-nominal characteristic{s} of the target. %{Based on \cite{miller1981inverse},}  
The inverse of $\bm{J}_\mathrm{c}$ can be computed as:
\be
\label{4}
\bm{J}_\mathrm{c}^{-1} = {\bm{J}}_{\mathrm{c}0}^{-1}+\tilde{\bm{J}_\mathrm{c}}^{*}
\ee
where $\tilde{\bm{J}}_\mathrm{c}^{*} = -(\bm{I}_3+\bm{J}_{\mathrm{c}0}^{-1}\tilde{\bm{J}}_\mathrm{c})^{-1}\bm{J}_{\mathrm{c}0}^{-1}\tilde{\bm{J}}_\mathrm{c}\bm{J}_{\mathrm{c}0}^{-1}$. 
% \bm{J}_\mathrm{c}\dot{\bm{\omega}} = -\bm{\omega}^{\times}\bm{J}_\mathrm{c}\bm{\omega}+\bm{u}+\bm{\tau}_\mathrm{d}

\YLnew{Thus, the attitude error dynamics of combined spacecraft with uncertainties can be \TR{fomulated as}
\be
\label{5}
\left\{
\begin{array}{ll}
\dot{\bm{q}}_\mathrm{e} &= \frac{1}{2}(q_{\mathrm{e}0}\bm{I}_{3}+\bm{q}_\mathrm{e}^{\times})\bm{\omega}\\
\dot{\bm{\omega}} &= -\bm{J}_{\mathrm{c}0}^{-1}\bm{\omega}^{\times}\bm{J}_{\mathrm{c}0}\bm{\omega}+\bm{J}_{\mathrm{c}0}^{-1}\bm{u}+\bm{J}_{\mathrm{c}0}^{-1}\bm{d}(\bm{\omega},\bm{u})
\end{array}\right.
\ee
where $\bm{d}(\bm{\omega},\bm{u})= -\bm{J}_{\mathrm{c}0}\tilde{\bm{J}}_\mathrm{c}^{*}(\bm{\omega}^{\times}\bm{J}_{\mathrm{c}0}\bm{\omega})-\bm{\omega}^{\times}\tilde{\bm{J}}_\mathrm{c}^{*}\bm{\omega}-\bm{J}_{\mathrm{c}0}\tilde{\bm{J}}_\mathrm{c}^{*}\bm{\omega}^{\times}\tilde{\bm{J}}_\mathrm{c}^{*}\bm{\omega}+ \bm{J}_{\mathrm{c}0}\tilde{\bm{J}}_\mathrm{c}^{*}\bm{u}+(\bm{I}_{3}+\bm{J}_{\mathrm{c}0}\tilde{\bm{J}}_\mathrm{c}^{*})\bm{\tau}_\mathrm{d}$. The dynamic model can be written in a compact form:}
\be
\label{6}
\dot{\bm{x}}=\underbrace{\bm{f}(\bm{x},\bm{u})}_{\YL{\text{nominal model}}}~+~\bm{B}\!\underbrace{\breve{\bm{\Delta}}(\bm{x},\bm{u})}_{\text{unknown model}}
\ee
where $\bm{x}=\mathrm{vec}(\bm{q}_\mathrm{e}, \bm{\omega}) \in \mathbb{R}^{6}$ is the state vector of the combined spacecraft,  $\bm{B}=[\mathbf{0}_{3\times 3}~ \bm{I}_3]^{\top}\in\mathbb{R}^{6\times 3}$ is a projection matrix, $\bm{f}(\bm{x},\bm{u})$ denotes the known, nominal part of the dynamics which has the following form:
\be
\label{7}
\bm{f}(\bm{x},\bm{u}) = 
\left[\begin{array}{c}\frac{1}{2}(q_{\mathrm{e}0}\bm{I}_{3}+\bm{q}_\mathrm{e}^{\times})\bm{\omega}\\
-\bm{J}_{\mathrm{c}0}^{-1}\bm{\omega}^{\times}\bm{J}_{\mathrm{c}0}\bm{\omega}+\bm{J}_{\mathrm{c}0}^{-1}\bm{u}\end{array}
\right]\in \mathbb{R}^{6}.
\ee
The \YL{state and control-dependent} unknown model $\breve{\bm{\Delta}}(\bm{x},\bm{u})=\bm{J}_{\mathrm{c}0}^{-1}\bm{d}(\bm{\omega},\bm{u})\in\mathbb{R}^3$ includes the model uncertainties and the {reactive} attitude maneuvering torque of the target. \LY{Note that $\breve{\bm{\Delta}}$ also implicitly depends on $t$ because it contains the time-varying external disturbance $\bm{\tau}_{\mathrm{d}}(t)$.} \YLnew{In order to {be able to} design {a} controller, the {system is required to satisfy the} following {non-restrictive conditions}:} %reasonable assumptions are made:}

\RV{\begin{condition} \label{cond:1}
{\rm \YLnew{There exists known and bounded constants $\lambda_J$, $\lambda_c>0$, such that the nominal part of the inertia matrix satisfies $\lambda_{\rm{max}}(\bm{J}_{\mathrm{c}0})\leq\lambda_J$ and $\lambda_{\rm{min}}(\bm{J}_{\mathrm{c}0})\geq\lambda_c$.}}
\end{condition}
}

\begin{condition} \label{cond:2}
{\rm  \YLnew{The unknown function  $\breve{\bm{\Delta}}$ is globally bounded.}}
\end{condition}

The combined spacecraft performs as a ``black box" {data-generating system}, i.e., only input and output data {is available from} %can be provided by
it, just in case of a real spacecraft.

\begin{assumption} \label{assmp:1}
{\rm \YLnew{Measurements of the state {(sampling of $\bm{x}(t)$)} are {obtained} %sampled 
online {under the} 
%in a synchronized manner with 
sampling time $T_\mathrm{s}\in\mathbb{R}^{+}$.  %that is,
{The sampling provided value of the states is denoted as} $\bm{q}_\mathrm{e}[k] := \bm{q}_\mathrm{e}(kT_\mathrm{s})$, $\bm{\omega}[k] := \bm{\omega}(kT_\mathrm{s})$, and $\bm{x}[k] := \bm{x}(kT_\mathrm{s})$. Also, we assume that an approximation of the state derivatives $\dot{\bm{x}}[k]$ can be obtained by numerical differentiation. %, which is a common setting in practical applications.
The approximation error can be considered as {being} part of {the} measurement noise.
Additionally, the control input $\bm{u}$ is generated via ideal \emph{zero-order-hold} (ZOH) actuation with no delay {and synchronized with the sampling}, which means that the discrete control signal will {be kept} constant until the next %time 
{sampling} moment:
\be
\bm{u}(t) := \bm{u}[k],~\forall t\in(kT_\mathrm{s},(k+1)T_\mathrm{s}].
\ee
}}
\end{assumption}

\vspace{-0.7cm}

\subsection{\YL{Control Objectives}}
\label{sub:sec:ctrlobj}
\YLnew{The primary goal of this paper is to design an adaptive online learning attitude takeover control strategy for the combined spacecraft in the presence of unknown dynamics and attitude maneuverability of the target, such that the attitude of the combined spacecraft {follows the} %maneuvers to a 
desired orientation $\bm{Q}_\mathrm{d}$, while the attitude error $\bm{Q}_\mathrm{e}$ and the angular velocity $\bm{\omega}$ are ultimately uniformly bounded, and converge to a small set containing the origin. }
% Specifically, the data-driven model should be capable of making full use of the
% online streaming data to update in real time with a low computational load.

% \YL{The control objective of this paper is to design a GP-based online learning  controller to ensure the combined spacecraft system states (\bmQ,\bmω)(\bm{Q},\bm{\omega}) converge to (\bm1,\bm0)(\bm{1},\bm{0}) in the presence of unmodeled dynamics and attitude maneuverability of target. Furthermore, the proposed recursive online sparse GP is employed to obtain a data-driven model of \bmΔ(\bmx,\bmu)\bm{\Delta}(\bm{x},\bm{u}), which is used for feed-forward compensation of the unknown function.}

\section{Gaussian Process for Online Model Learning} \label{sec:3}
\YLnew{\RV{To achieve the aforementioned control objectives, %a model %${\bm{\Delta}}$ %(\tilde{\bm{x}})$ 
%of 
%$\breve{\bm{\Delta}}$ 
%{is aimed to be} 
%should be 
%identified {as the} %reliably because the dynamics given in  is a 
%simplified model \eqref{5} %which 
%is 
%{insufficient} %not enough 
%to describe the attitude motion {dynamics} of the combined spacecraft accurately.} % in complex scenarios. 
%{Hence, in} 
in}
this section, our goal is to construct a data-driven and probabilistic model ${\bm{\Delta}}$ {of the unknown dynamics $\breve{\bm{\Delta}}$} from previously collected measurement data, and improve the accuracy of the regressed model gradually as more data becomes available \TR{and track possible time variation of $\breve{\bm{\Delta}}$}.} \vspace{-0.3cm}

\subsection{Gaussian Process Regression} 
\label{sec:sub:fullGP}

{GP \emph{regression} (GPR)} is a powerful non-parametric framework for {learning} nonlinear {functions from data}, {where} the GP itself can be {seen} %described 
as a distribution over functions \cite{rasmussen2003gaussian}. The main advantage of {GPR} is that it not only provides the estimated mean of the unknown function, but also \TR{its} variance, which implies the regression accuracy, namely, the model confidence. \YLnew{To approximate the unknown function $\breve{\bm{\Delta}}$, we {consider} $\bm{\Delta}(\tilde{\bm{x}})$ as a GP which is trained based on the following data set consisting of $N$ collected sampled measurements:}
\be
\label{8}
\mathcal{D}_N=\{\yi,~\tildexi\}_{i=1}^{N}
\ee
where the state-input pairs  $\tildexk=\mathrm{vec}(\xk,\uk)\in\mathbb{R}^9$ and $\yk=\breve{\bm{\Delta}}(\tildexk)+\eek\triangleq\dot{\bm{x}}[k]-\bm{f}(\xk),\uk)+\eek$
 denote the training inputs and outputs, respectively, and $\eek$ represents the \emph{independent and identically distributed} (i.i.d.) measurement noise with $\eek\sim\mathcal{N}(\mathbf{0},{\sigma}^{2}_\mathrm{\epsilon}\bm{I})$ and additional approximation error result{ing} from numerical differentiation. 
%  Due to measurement noise and additional effects
% \be
% \label{9}
% \yk=\bm{g}(\tildexk)+\eek
% \ee

\begin{figure}[t]
  \centering\includegraphics[width= 2.6 in]{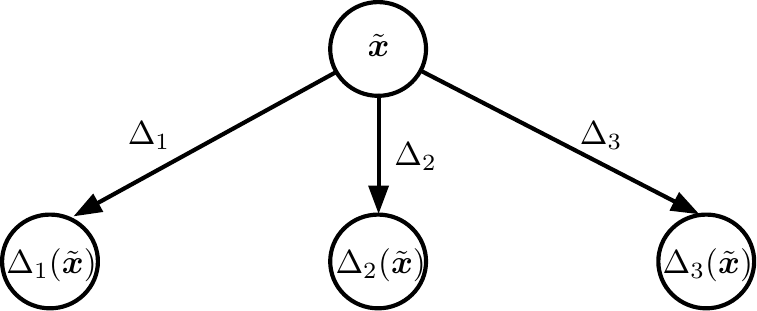} \vspace{2mm}
  \caption{{I}llustration of {the} multi-dimensional GP structure.} \vspace{-4mm}
  \label{fig:2}
\end{figure}

% where the state-input pairs \bm˜x\bm{\tilde{x}} and 
% \be
% \label{10}
% \bm{\Delta}(\bm{\tilde{x}}) = \bm{B}^{\dagger} (\dot{\bm{x}}-\bm{f}(\bm{x},\bm{u}))
% \ee
% are the training inputs and outputs, respectively, and (⋅)†(\cdot)^{\dagger} denotes the pseudo inverse. \YLnew{Particularly, we assume that an approximation of the state derivative ˙\bmx(t)\dot{\bm{x}}(t) is  
% available (e.g. obtained by numerical differentiation: ˙\bmx(t)=(\bmx(k)−\bmx(k−1))/Ts\dot{\bm{x}}(t)=(\xk-\bm{x}(k-1))/T_{\rm{s}}). Furthermore, the approximation error can be considered as a part of measurement noise.} 
%Then, a
{A} vectorial \emph{Gaussian process} $\mathcal{GP}:\mathbb{R}^9\to\mathbb{R}^3$ assigns to every point $\bm{\tilde{x}}\in\mathbb{R}^9$ a random variable $\mathcal{GP}(\bm{\tilde{x}})$ taking values in $\mathbb{R}^{3}$ such that, for any finite set $\{\tildext\}_{\tau=1}^N \subset \mathbb{R}^{3}$, the joint probability distribution of $\mathcal{GP}\left(\tildexfirst\right), \ldots, \mathcal{GP}\left(\tildexN\right)$ is multi-dimensional Gaussian. This constitutes a prior distribution over functions, which is denoted by:
\be
\label{11}
\bm{\Delta}(\bm{\tilde{x}}) \sim \mathcal{GP}(\bm{\mu}(\bm{\tilde{x}}),~\bm{\kappa}(\bm{\tilde{x}},\bm{\tilde{x}}')),
\ee
where $\bm{\mu}(\bm{\tilde{x}})$ is the mean function and $\bm{\kappa}\left(\tilde{\bm{x}}, \tilde{\bm{x}}^{\prime}\right)\triangleq\operatorname{cov}\left(\bm{\Delta}(\tilde{\bm{x}}), \bm{\Delta}\left(\tilde{\bm{x}}^{\prime}\right)\right)$ is the positive semi-definite covariance function which {corresponds to} a measure {of} %the 
correlation of any two data points $(\tilde{\bm{x}},\tilde{\bm{x}}^{\prime})$.
Furthermore, the GP model is usually implemented for each dimension of the GP output separately, i.e. in terms of scalar-valued ${\Delta}_j(\bm{\tilde{x}}) \sim \mathcal{GP}({\mu}_j(\bm{\tilde{x}}),~{\kappa}_j(\bm{\tilde{x}},\bm{\tilde{x}}'))$ which approximates the corresponding $\breve{\Delta}_j$ with ${j\in\mathbb{I}_1^3}$. %=1,2,3$. 
\RV{The structure of the multi-dimensional GP is illustrated in Fig.~\ref{fig:2}, where the outputs are assumed to be uncorrelated, i.e., $\breve{\Delta}_j$ is mutually independent, \TR{resulting in the choice of $\bm{\kappa}=\mathrm{diag}(\kappa_1,\kappa_2,\kappa_3)$.} It should be noted that \TR{this} assumption is commonly used in GP-based control approaches
and \TR{it is} mild for a real spacecraft task.
%Thus, it inherently results in the choice of $\bm{\kappa}=\mathrm{diag}(\kappa_1,\kappa_2,\kappa_3)$. 
}\YLnew{In this paper, the kernel function is chosen from the exponential family, i.e., the \emph{Squared Exponential Automatic Relevance Determination} (SEARD) \TR{is taken} as a prior due to its universal approximation capability:}
\be
\label{12}
\kappa_j\left(\bm{\tilde{x}}, \bm{\tilde{x}}'\right)\!=\!\sigma_{{\rm{f}},j}^{2} \exp \left(-\frac{1}{2}(\bm{\tilde{x}}- \bm{\tilde{x}}')^{\top}\bm{\Lambda}_{j}^{-1}(\bm{\tilde{x}}- \bm{\tilde{x}}')\right)
\ee
where the diagonal matrix $\bm{\Lambda}_{j}=\mathrm{diag}(\lambda_{j,1}^2,\ldots,\lambda_{j,9}^2)$ and $\sigma_{{\rm{f}},j}^{2}\in\mathbb{R}^{+}$ are the length-scale hyperparameters and signal variance, respectively. \YLnew{In order to {be able to capture the unknown function $\breve{\bm{\Delta}}$ using the chosen kernel defined GP \eqref{11}, the following condition is required to be satisfied:}}  %design the controller, the following reasonable assumptions are made:}

\YLnew{\begin{condition} \label{assump:4}
{\rm %The function  
{Each} ${{\breve{{\Delta}}}_j}$ has a bounded \emph{reproducing kernel Hilbert Space} (RKHS) norm w.r.t. the chosen %SEARD 
kernel ${\kappa}_{j}(\bm{\tilde{x}},\bm{\tilde{x}}')$, that is, $\|{{\breve{{\Delta}}}_j}\|_{\mathcal{H}}<\infty$.}
\end{condition}}

\YLnew{\begin{remark} 
{\rm The RKHS norm of ${\breve{{\Delta}}}_j$ can be interpreted as a quantitative assessment of the function smoothness, {indicating} that the function is ``well-behaved" w.r.t. the selected kernel.} %in GP.}
\end{remark}}

According to the training data set \eqref{8}, we define $\tilde{\bm{X}}={\rm{vec}}(\tildexfirst, \ldots, \tildexN)$ and $\bm{Y}_j={\rm{vec}}\left(\yjfirst, \ldots, \yjN \right)$ for \YLnew{$j\in\mathbb{I}_1^3$}. The Gaussian prior for the function $\bm{\Delta}$ and the model likelihood of $\mathcal{D}_N$ is denoted as:
\begin{subequations}
\begin{align}
\label{14}
\mathbb{P}(\bm{\Delta}_{j})&=\mathcal{N}(\bm{\Delta}_{j}  \!\mid\! \bf{0}, \bm{K}_{N,j})\\
\label{14b}
\mathbb{P}(\bm{Y}_{j} \!\mid\! \bm{\Delta}_{j})&=\mathcal{N}\left(\bm{Y}_{j} \!\mid\! \bm{\Delta}_{j}, \sigma^{2}_{\epsilon,j}\bm{I}_N\right)
\end{align}
\end{subequations}
where the so-called Gram matrix $\bm{K}_{N,j}\in\mathbb{S}^{N\times N}$ represents the symmetric and semi-definite covariance matrix on the training set $\mathcal{D}_N$:
\be
\label{15}
\bm{K}_{N,j}\!=\!\!\left[\!\begin{array}{ccc}
\kappa_{j}\left(\tildexfirst, \tildexfirst\right) & \cdots & \kappa_{j}\left(\tildexfirst, \tildexN\right) \\
\vdots & \ddots & \vdots \\
\kappa_{j}\left(\tildexN, \tildexfirst\right) & \cdots & \kappa_{j}\left(\tildexN, \tildexN\right)
\end{array}\!\right]
\ee
and $\bm{\Delta}_j={\rm{vec}}(\Delta_j(\tildexfirst), \ldots, \Delta_j(\tildexN)$.

According to Bayes' Theorem, the posterior distribution can be obtained by \emph{maximum a posterior} (MAP) estimation:
\begin{multline}
\label{17}
\mathbb{P}(\bm{\Delta}_{j} \!\mid\! \bm{Y}_j)= \frac{\mathbb{P}(\bm{Y}_{j} \!\mid\! \bm{\Delta}_{j}) \mathbb{P}(\bm{\Delta}_{j})}{\mathbb{P}(\bm{Y}_{j} )}\\
\propto \mathcal{N}\bigr(\bm{\Delta}_{j} \!\mid\! \bm{K}_{N,j}\left(\bm{K}_{N,j}+\sigma^{2}_{\epsilon,j}\bm{I}_N\right)^{-1}\bm{Y}_{j}, \\
\sigma^{2}_{\epsilon,j}\bm{K}_{N,j}\left(\bm{K}_{N,j}+\sigma^{2}_{\epsilon,j}\bm{I}_N\right)^{-1}\bigr).
\end{multline}

Then, the predictive distribution of %the $j$-th dimension of function 
$\Delta_j({\bm{\tilde{x}}}^{*})$ on a test point ${\bm{\tilde{x}}}^{*}$ can be derived as: 
\be
\label{18}
\begin{aligned}
\mathbb{P}(\Delta_j^* &\!\mid\! \tilde{\bm{x}}^*, \mathcal{D}_N)\\
&=\int \mathbb{P}\left(\Delta_j^* \!\mid\! \tilde{\bm{x}}^*, \bm{\Delta}_{j}, \tilde{\bm{X}}\right) \mathbb{P}(\bm{\Delta}_{j} \!\mid\! \bm{Y}_j) \mathrm{d} \bm{\Delta}_j\\
&=\mathcal{N}(\mu_{{\Delta},{j}}(\bm{\tilde{x}}^{*}),\sigma^{2}_{{\Delta},{j}}(\bm{\tilde{x}}^{*}))
\end{aligned}\vspace{0.1cm}
\ee
where the first term inside the integral satisfies the joint distribution: \vspace{0.1cm}
\begin{align}
&\mathbb{P}\left(\Delta_j^* \!\mid\! \tilde{\bm{x}}^*, \bm{\Delta}_{j}, \tilde{\bm{X}}\right)  \label{19} \\ \notag
&=\mathcal{N}\left(\Delta_j^*  \!\mid\! \bm{K}_{*N,j} \bm{K}_{N,j}^{-1}\bm{\Delta}_{j}, k_{**,j}-\bm{K}_{*N,j} \bm{K}_{N,j}^{-1} \bm{K}_{N*,j}\right). \vspace{0.1cm}
\end{align}
\RV{Combining ~\eqref{17}, \eqref{18} and \eqref{19}, we can obtain the posterior mean and variance function for each dimension $j$:}
\begin{subequations}
\begin{align}
\label{21}
\mu_{{\Delta},{j}}(\bm{\tilde{x}}^{*})&=\bm{K}_{*N,j}(\bm{K}_{N}^j+\sigma_{\epsilon,j}^2\bm{I})^{-1}\bm{Y}_{j},\\
\label{22}
\sigma^{2}_{{\Delta},{j}}(\bm{\tilde{x}}^{*})&={k}_{**,j}\!-\!\bm{K}_{*N,j}(\bm{K}_{N,j}\!+\!\sigma_{\epsilon,j}^2\bm{I})^{-1}\!\bm{K}_{N*,j},
\end{align}
\end{subequations}
which gives {a predictive} vectorial GP {with} %that the predictive 
mean and variance: % can be written in the compact form: 
\begin{subequations}
\label{23}
\begin{align}
\bm{\mu}_{\Delta}(\bm{\tilde{x}}^{*})&={\rm{vec}}(\mu_{{\Delta},{1}}(\bm{\tilde{x}}^{*}),\mu_{{\Delta},{2}}(\bm{\tilde{x}}^{*}),\mu_{{\Delta},{3}}(\bm{\tilde{x}}^{*})),\\
\bm{\Sigma}_{\Delta}(\bm{\tilde{x}}^{*}) &= \mathrm{diag}(\sigma^{2}_{{\Delta},{1}}(\bm{\tilde{x}}^{*}), \sigma^{2}_{{\Delta},{2}}(\bm{\tilde{x}}^{*}), \sigma^{2}_{{\Delta},{3}}(\bm{\tilde{x}}^{*})).
\end{align}
\end{subequations}

Furthermore, to obtain an optimal choice of the hyperparameters\footnote{\RV{We initialize $\bm{\theta}_{j}$ based on $\mathcal{D}_N$: $\mathrm{log}\bm{\theta}_j^0=\mathrm{vec}(\mathrm{log}(\mathrm{std}(\bm{Y}_j/10)),$ $\mathrm{log}(\mathrm{std}(\bm{Y}_j)),\mathrm{log}(\mathrm{std}(\tilde{\bm{X}})))$. In \cite{chen2018priors}, it was illustrated that although the initial guess for the hyperparameters may have influences on the \TR{optimization} results, its impact on the regression accuracy of the GP model is almost negligible. Therefore, it is generally advisable to choose a relatively simple initial guess.} }$\bm{\theta}_{j}=\mathrm{vec}(\sigma_{\epsilon,j}^{2},\sigma_{{\rm{f}},j}^{2},\lambda_{j,1},\ldots,\lambda_{j,9})$, the GP model \eqref{11} is trained by maximizing the log-likelihood $\mathcal{L}(\bm{\theta}_{j})=\log \mathcal{N}\left(\mathbf{0},~\bm{K}_{N,j} + \sigma^{2}_{\epsilon,j}\bm{I}_N\right)$ w.r.t. $\bm{\theta}_{j}$  for each dimension $j$:
\be
\label{25}
\hat{\bm{\theta}}_{j} \in \arg \max _{\bm{\theta}_{j}} \mathcal{L}(\bm{\theta}_{j}),
\ee
which optimization problem can be efficiently solved by {a} conjugate gradient-based algorithm \cite{burden2015numerical}.\vspace{0.2cm}

\subsection{Recursive Online Sparse GP Regression} \label{sec:3:b}% rename

As we could see in \eqref{25}, the computational complexity for the conjugate gradient-based algorithm is $\mathcal{O}(N^3)$ per iteration, which is cubic in terms of the size of the training data set. According to \eqref{21} and \eqref{22}, the computational complexity for predictive mean and variance per test case is $\mathcal{O}(N)$ and $\mathcal{O}(N^2)$, respectively. Thus, the standard GP is not suitable for large training data sets. However, it is essential to collect a large number of data pairs to explore the state space as much as possible and ensure a sufficiently high regression accuracy. Furthermore, the GP model presented in Section III.\ref{sec:sub:fullGP}
is an offline modeling approach, i.e., the trained GP model is kept fixed online and considered to be sufficient to describe the various on-orbit scenarios. Nevertheless, for the combined spacecraft takeover control missions, the unknown {part of the dynamics} ${\breve{\bm{\Delta}}}$ may be time-varying due to further attitude maneuvers of the target. Therefore, this section proposes a recursive online sparse GP algorithm (denoted by ROSGP), which significantly reduces the computational complexity of the data-driven model learning while making full use of the online streaming data to update the GP model in real time.
\subsubsection{Sparse GP with inducing points}
As discussed in Section III.\ref{sec:sub:fullGP}, the diagonal assumption for the prior covariance function $\bm{\kappa}$ allows for the training of the predictive distribution independently for each dimension $j$. Thus, for the simplicity of notation, we will drop the indexing for the output dimension $j$.

\RV{We \TR{will introduce a novel} generalization of \TR{the original} \emph{sparse GP with inducing inputs} (SPGP) %, \TR{originally} which \TR{has been} first introduced in 
\cite{snelson2005sparse}} \TR{approach in terms of an efficient online update step. For this we first briefly summarize the SPGP method}. The main idea of sparse GP is to find a set of inducing inputs  $\bm{\tilde{X}}_\mathrm{u}=\mathrm{vec}(\bm{\tilde{x}}_{\mathrm{u},1},...,\bm{\tilde{x}}_{\mathrm{u},M})$ corresponding to inducing outputs $\bm{\Delta}_\mathrm{u}=\!\mathrm{vec}({\Delta}_{\mathrm{u},1},...,{\Delta}_{\mathrm{u},M})$ of size $M\ll N$. Akins to \eqref{14}, the inducing points follow the Gaussian prior distribution:
\be
\label{26}
\mathbb{P}(\bm{\Delta}_\mathrm{u} )=\mathcal{N}(\bm{\Delta}_\mathrm{u}\!\mid\! \mathbf{0}, \bm{K}_M)
\ee
where $\bm{K}_M\in\mathbb{S}^{M\times M}$ denotes the Gram matrix in terms of inducing inputs. %If the prior distribution \eqref{26} is selected appropriately, $\mathcal{D}_M$ should have the same distribution as the full data set $\mathcal{D}_N$.
According to the \emph{fully independent training conditional approximation} (FITC), given the GP inputs {with \eqref{26}}, the function value{s} $\bm{\Delta}={\rm{vec}}(\Delta(\tildexfirst), \ldots, \Delta(\tildexN)$ \TR{are} %satisfies the 
i.i.d.
%condition. 
Then, the model likelihood is \vspace{-2mm} %induced as:
\begin{align}
\label{28}\mathbb{P}(\bm{Y}\!\mid\!\tilde{\bm{X}},\bm{\Delta}_\mathrm{u}, \bm{\tilde{X}}_\mathrm{u})&= \prod_{i=1}^{N}\mathbb{P}(\yi \!\mid\!\tildexi,\bm{\Delta}_\mathrm{u}, \bm{\tilde{X}}_\mathrm{u})\\ \notag
&=\mathcal{N}(\bm{Y}\!\mid\!\bm{K}_{NM}\bm{K}_{M}^{-1}\bm{\Delta}_\mathrm{u},\bm{\Gamma}+\sigma_{\epsilon}^2\bm{I}_N)
\end{align}
where $[\bm{K}_{MN}]_{i,j}={\kappa}({\bm{\tilde{x}}_{\mathrm{u},i}},\tildexj)$ denotes the covariance matrix between $\mathcal{D}_M$ and $\mathcal{D}_N$, $\bm{Q}_N=\bm{K}_{NM} \bm{K}_{M}^{-1} \bm{K}_{M N}$ can be seen as an approximation of $\bm{K}_N$, \TR{and} $\bm{\Gamma}=\mathrm{diag}\left(\bm{K}_N-\bm{Q}_N\right) \in \mathbb{R}^{N \times N}$ represents the diagonal covariance matrix which obtains its diagonal structure from the independence between $\bm{\Delta}$ and $\bm{\Delta}_\mathrm{u}$.

On the basis of {the} Bayes's Theorem {and} the Gaussian prior \eqref{26}, the approximated posterior distribution is {given} by {the} MAP estimat{e}:
\be
\label{29}
\begin{aligned}
\mathbb{P}(\bm{\Delta}_\mathrm{u} &\!\mid\! \bm{Y}, \bm{\tilde{X}}_\mathrm{u}, \tilde{\bm{X}})\propto\mathbb{P}(\bm{Y} \!\mid\! \bm{\Delta}_\mathrm{u}, \bm{\tilde{X}}_\mathrm{u}, \tilde{\bm{X}}) \mathbb{P}(\bm{\Delta}_\mathrm{u}) \\
&=\mathcal{N}\left(\bm{\Delta}_\mathrm{u} \!\mid\! \bm{K}_{NM}\bm{Q}_{M}^{-1} \bm{K}_{MN}(\bm{\Gamma}+\sigma_{\epsilon}^{2} \bm{I}_N\right)^{-1} \bm{Y},\\ 
&~~~~~~~~~~~~~~~~\qquad\qquad\quad\bm{K}_{M}\bm{Q}_{M}^{-1}\bm{K}_{M})
\end{aligned}
\ee
where $\bm{Q}_{M}=\bm{K}_{M}+\bm{K}_{MN}(\bm{\Gamma}+\sigma_{\epsilon}^{2} \bm{I}_N)^{-1}\bm{K}_{NM}$.

%% dimension 
Consequently, the associated posterior distribution of  ${\Delta}({\bm{\tilde{x}}}^{*})$  at a new test point $\bm{\tilde{x}}^{*}$ is computed by
\begin{align}
 \mathbb{P}(\Delta^{*}|&\mathcal{D}_N, \bm{\tilde{x}}^{*},\bm{\tilde{X}}_\mathrm{u}) 
 % \notag \\ 
% \notag
% &=\int \mathbb{P}(\Delta^{*}|\bm{\Delta}_\mathrm{u},\bm{\tilde{X}}_\mathrm{u},\bm{\tilde{x}}^{*})\mathbb{P}(\bm{\Delta}_\mathrm{u} \!\mid\! \bm{Y}, \bm{\tilde{X}}_\mathrm{u}, \tilde{\bm{X}})\mathrm{d}\bm{\Delta}_\mathrm{u}\\
=\mathcal{N}(\mu_{\Delta}(\bm{\tilde{x}}^{*}),\sigma^{2}_{{\Delta}}(\bm{\tilde{x}}^{*})) \label{30}
\end{align}
with predictive mean and variance as:
\begin{subequations}
\begin{align}
\label{31}
\mu_{{\Delta}}(\bm{\tilde{x}}^{*})&=\bm{K}_{*M}\bm{Q}_{M}^{-1} \bm{K}_{MN}\left(\bm{\Gamma}+\sigma_{\epsilon}^{2} \bm{I}_N\right)^{-1} \bm{Y},\\
\label{32}
\sigma^{2}_{{\Delta}}(\bm{\tilde{x}}^{*})&=k_{**}-\bm{K}_{*M}\left(\bm{K}_{M}^{-1}-\bm{Q}_{M}^{-1}\right) \bm{K}_{M*}.
\end{align}
\end{subequations}

\RV{ \TR{The inducing data points $\mathcal{D}_M=\{\bm{\Delta}_{\mathrm{u}},\tilde{\bm{X}}_{\mathrm{u}}\}$ can be seen as additional hyper-parameters and are}
optimized along with \TR{the original} GP hyperparameters $\bm{\theta}=\operatorname{vec}\left(\sigma_{\epsilon}^2, \sigma_{\mathrm{f}}^2, \lambda_{1}, \ldots, \lambda_{9}\right)$ by maximizing the marginal likelihood function, which can be computed by integrating \eqref{26} and \eqref{28}:
\be
\label{spgp_lik}
\begin{aligned}
\mathbb{P}(\bm{Y})&= \int \mathbb{P}(\bm{Y}\!\mid\!\tilde{\bm{X}},\bm{\Delta}_\mathrm{u}, \bm{\tilde{X}}_\mathrm{u})\mathbb{P}(\bm{\Delta}_\mathrm{u} )\mathrm{d}\bm{\Delta}_\mathrm{u}\\
& = \mathcal{N}\left(\mathbf{0},\bm{Q}_N + \bm{\Gamma}+\sigma_{\epsilon}^2\bm{I}_N\right).
\end{aligned}
\ee
%Thus, in SPGP method, the locations of inducing inputs serve as additional hyperparameters. 
\TR{To initialize the optimization, one reliable approach is to pick random points from the original data set $\mathcal{D}_N$.} %Furthermore, the inducing data set $\mathcal{D}_M=\{\bm{\Delta}_{\mathrm{u}},\tilde{\bm{X}}_{\mathrm{u}}\}$ is initialized randomly inside $\mathcal{D}_N$ as the initial guess, and subsequently 
}

It is worth mentioning that the inverse of $\bm{K}_{N}+\sigma^{2}_{\epsilon}\bm{I}_N$ is reduced to the inverse of the diagonal matrix $\bm{\Gamma}+\sigma_{\epsilon}^{2}\bm{I}_N$. In the SPGP based on the FITC assumption, the computational load $\mathcal{O}(M^2N)$ mainly comes from the matrix multiplication $\bm{K}_{MN}(\bm{\Gamma}+\sigma_{\epsilon}^{2} \bm{I}_N)^{-1}\bm{K}_{NM}$. Subsequently, for each test point $\bm{\tilde{x}}^{*}$, the computational complexity for corresponding predictive mean and variance is decreased to $\mathcal{O}(M)$ and $\mathcal{O}(M^2)$, respectively. Moreover, compared to the marginal likelihood of the standard GP, $\bm{Q}_N$ is a low-rank approximation of Gram matrix $\bm{K}_N$, which reduces the computational complexity from $\mathcal{O}(N^3)$ to $\mathcal{O}(M^2N)$ during the hyperparameter training. 
%In addition, the inducing data set $\mathcal{D}_{M}=\left\{\tilde{\bm{X}}_{\mathrm{u}}, \bm{\Delta}_{\mathrm{u}}\right\}$ can be regarded as a part of the hyperparameters to be optimized. 
Similar to the standard GP, the SPGP model can be trained by maximizing the log form of the likelihood \eqref{spgp_lik} to find the optimal hyperparameters and $\mathcal{D}_{M}$ by means of {a} conjugate gradient-based algorithm, which has a computational complexity of $\mathcal{O}(M^2N+3MN)$ per iteration.

\subsubsection{Recursive Online Sparse GP Regression Algorithm}
In this subsection, a novel online update strategy for SPGP is proposed, where the offline trained GP model is recursively updated with the online measurement data sampled at the current time moment {$k\in\mathbb{N}$}. 

\RV{Equation \eqref{31} can be rewritten into a linear combination of $M$ kernel functions with the current time moment $k$:}
\be
\label{34}
{\mu}_{\Delta,k}(\bm{\tilde{x}})=\sum_{j=1}^{M}\alpha_{j}\kappa(\tilde{\bm{x}}_{\mathrm{u},j},\tildexk)= \bm{\alpha}^{\!\top}\![k] \bm{K}_{M{[k]}}
\ee
where $\bm{\alpha}[0]=\bm{Q}_{M}^{-1} \bm{K}_{MN}\left(\bm{\Gamma}+\sigma_{\epsilon}^{2} \bm{I}_N\right)^{-1} \bm{Y}\in\mathbb{R}^{M}$ is the initial weight vector obtained from offline training {and} $ [\bm{K}_{M{[k]}}]_j = \kappa(\tilde{\bm{x}}_{\mathrm{u},j},\tildexk)$ is the corresponding kernel \TR{slice evaluated at} %function vector of 
the input $\bm{\tilde{x}}[k]$.

At time moment $k$, consider the data set $\mathcal{D}_{k}$ given in \eqref{8}. Define a performance index $\mathcal{W}(\bm{\alpha})$ w.r.t. $\bm{\alpha}$ over the extended data set $\mathcal{D}_{k}=\mathcal{D}_{k-1}\cup{({y}[k],\bm{\tilde{x}}[k])}$ which contains $N_k$ data pairs:
\be
\label{35}
\mathcal{W}(\bm{\alpha})=\sum_{i=1}^{N_k} \lambda^{N_k-i}\left(y[i]-\bm{\alpha}^{\top} \bm{K}_{Mi}\right)^{2}+\varsigma \lambda^{N_k+1}\|\bm{\alpha}\|^{2}
\ee
where $[\bm{K}_{Mi}]_j= \kappa(\tilde{\bm{x}}_{\mathrm{u},j},\tildexi)$, $0<\lambda\leq1$ is a user-defined parameter, also known as the forgetting factor. The choice of parameter $\varsigma$ will be discussed later in this section. 
The weight vector $\alphak$ at \TR{step} $k$ %-th step should 
\TR{is taken as the minimizer of}
%minimize 
\eqref{35}, i.e.:
\be
\label{36}
\alphak =\arg \min_{\bm{\alpha}\in \mathbb{R}^M} \mathcal{W}(\bm{\alpha}).
\ee

The optimization problem \eqref{36} has an analytical solution. \TR{Note that}
\be
\label{37}
\Phik\alphak  \rhok
\ee
where
\begin{subequations}
\begin{align}
\label{38}
\Phik&=\sum_{i=1}^{N_k} \lambda^{N_k-i} \bm{K}_{Mi} \bm{K}_{Mi}^{\top}+\varsigma\lambda^{N_k+1} \bm{I}_M,\\
\label{39}
\rhok&=\sum_{i=1}^{N_k} \lambda^{N_k-i} \bm{K}_{Mi} y[i].
\end{align}
\end{subequations}

Next, \eqref{38} and \eqref{39} can be %further 
written in a separate form:
\begin{subequations}
\begin{align}
\label{40}
\Phik&=\lambda\Phikfirst+\bm{K}_{M{[k]}}\bm{K}_{M{[k]}}^{\top},\\
\label{41}
\rhok&=\lambda\rhokfirst+\bm{K}_{M{[k]}}{y}[k].
\end{align}
\end{subequations}

Using formula Woodbury’s matrix inverse for \eqref{40}, one has:
\be
\label{42}
\bm{\Phi}^{-1}[k] =  \lambda^{-1}\bm{\Phi}^{-1}[k-1] - \lambda^{-1}\Lk\bm{K}_{M{[k]}}^{\top}\bm{\Phi}^{-1}[k-1]
\ee
where
\be
\label{43}
\Lk=\frac{\lambda^{-1} \bm{\Phi}^{-1}[k-1] \bm{K}_{M{[k]}}}{1+\lambda^{-1} \bm{K}_{M{[k]}}^{\top} \bm{\Phi}^{-1}[k-1]\bm{K}_{M{[k]}}}.
\ee
For the convenience of notation, we define $\Pk=\bm{\Phi}^{-1}[k]$. Thus, from \eqref{42} and \eqref{43}, we have:
\begin{subequations}
\begin{align}
\label{44}
\Pk &=  \lambda^{-1}\Pkfirst - \lambda^{-1}\Lk\bm{K}_{M{[k]}}^{\top}\Pkfirst,\\
\label{44-1}
\Lk&=\Pk\bm{K}_{M{[k]}}.
\end{align}
\end{subequations}

Subsequently, combining \eqref{37}, \eqref{41}, \eqref{42}, and \eqref{43}, one can derive:
\begin{equation}
\label{45}
\alphak=\alphakfirst-\Lk \bm{K}_{M{[k]}}^{\top} \alphakfirst+\Lk {y}[k]
\end{equation}

Finally, the weight vector $\alphak$ can be recursively updated by:
\be
\label{46}
\alphak = \alphakfirst + \Lk {r}[k]
\ee
where ${r}[k]= {y}[k]-\bm{\alpha}^{\top}_{j}[k-1]\bm{K}_{M{[k]}} $. 
% The recursion starts from \bmα0\bm{\alpha}[0] and \bmP0\bm{P}[0], which can be seen as a prior of the GP.

\begin{remark}
{\rm The online update routine starts from the initial weight vector $\bm{\alpha}[0]$ and the initial user-defined matrix $\bm{P}[0]$. \RV{$\bm{\alpha}[0]$ can be 
\TR{obtained by offline training of the}
%provided by an offline trained 
GP and $\bm{P}[0]$ is usually selected as $\bm{P}[0]=\varsigma^{-1}\bm{I}_M$ with $0<\varsigma\leq 1$.} The choice of $\varsigma$ is on the basis of the confidence level of the offline trained GP.}
\end{remark}
%\TR{Generally}, a smaller $\varsigma$ also leads to a faster convergence rate. }

\begin{remark} 
{\rm It is worth mentioning that, the convergence of the proposed recursive online sparse GP is inherently guaranteed because it minimizes the performance index \eqref{35} at each iterative step.}
\end{remark}

\begin{remark} 
{\rm \RV{
Compared with the \TR{existing} online GP methods, such as SOGP \cite{csato2002sparse,chowdhary2014bayesian}, and the evolving GP in \cite{kocijan2016modelling,maiworm2021online}, the proposed recursive online sparse GP do not involve any re-optimization of the hyperparameters \TR{nor it requires} data-dictionary update at every time step, which ensures low online computational cost.}
}
\end{remark}

\vspace{-1mm}
\section{GP-based Online Learning Control}
\label{sec:GPctrl}
\YLnew{Recalling the control objective given in Section II. \ref{sub:sec:ctrlobj}, the goal of this paper is to design a GP-based online learning control strategy to ensure that, {for a given attitude set point $\bm{Q}_\mathrm{d}=[q_{\mathrm{d}0}~\bm{q}_\mathrm{d}^\top]^\top\in{\mathbb{Q}}^3$,} the closed-loop {error and velocity} states $(\bm{Q}_\mathrm{e},\bm{\omega})$ converge to $(\bm{1},\bm{0})$ even %with
\TR{under} the unknown dynamics and attitude maneuverability of target. Furthermore, the proposed ROSGP algorithm is employed to derive a probabilistic model of $\breve{\bm{\Delta}}(\tilde{\bm{x}})$, which is utilized for feedforward compensation of the unknown function.}
 
\subsection{Controller Design}

In order to design the controller, we begin by considering the following lemmas and conditions.

\begin{lemma}\label{lemma:1}
{\rm \YLnew{Consider a GP {trained on $\mathcal{D}_N$ \TR{collected} from system \eqref{6}, which} satisfies {Condition} \ref{assump:4}. The estimation error $\|\bm{\mu}_{\Delta}(\bm{\tilde{x}})-{\breve{\bm{\Delta}}}(\bm{\tilde{x}})\|$ is bounded for all $\bm{\tilde{x}}$ on the compact set $\Omega \subset \mathbb{R}^{9}$ with probability $(1-\delta)^3$:}
\be
\label{47}
\begin{aligned}
\mathbb{P}\{\forall \bm{\tilde{x}} \!\in \!\Omega,~\|\bm{\mu}_{\Delta}(\bm{\tilde{x}})\!-\!{\breve{\bm{\Delta}}}(\bm{\tilde{x}})\|\leq\|\bm{\beta}\|\|\bm{\Sigma}_{\Delta}^{1/2}(\bm{\tilde{x}})\|\}\geq (1-\delta)^3
\end{aligned}
\ee
where $\delta\in(0,1)$, $\bm{\beta}=\mathrm{vec}(\beta_1, \beta_2, \beta_3)$ denotes:
\vspace{-1mm}
\be
\beta_{j} = \sqrt{2\|{\breve{\Delta}}_j\|_{\mathcal{H}}^{2}+300\gamma_{j}\log^{3}((N+1)/{\delta})}, 
\ee
and $\gamma_{j}=\max\frac{1}{2}\log|{\bm{I}_N+\sigma_{\epsilon,j}^{-2}\bm{K}_{N,j}}|$ represents the maximum information gain w.r.t. the kernel $\kappa_j, \forall j\in\mathbb{I}_1^3$.}
\end{lemma}
\vspace{-1mm}
\begin{proof} 
 See \cite[Lemma~2]{umlauft2018uncertainty}. This lemma is a vectorial generalization of the scalar case given in \cite{srinivas2012information}.
\end{proof}

\begin{proposition}\label{prop1}
{\rm {Consider a GP {recursively trained on the initial data set $\mathcal{D}_N$ and online samples obtained during time interval $[0,k]$ from system \eqref{6}  which satisfies  Condition \ref{assump:4}.} 
%\YLnew{Consider the system \eqref{6} and a trained GP which satisfies Assumption 4. Then 
{The} estimation error $\|\bm{\mu}_{\Delta,k}(\bm{\tilde{x}})-{\breve{\bm{\Delta}}}(\bm{\tilde{x}})\|$ is bounded for all $\bm{\tilde{x}}$ on the compact set $\Omega \subset \mathbb{R}^{9}$ with probability $(1-\delta)^3$:}
\be
\label{47}
\begin{aligned}
\mathbb{P}\{\forall \bm{\tilde{x}} \!\in \!\Omega,~\|\bm{\mu}_{\Delta,k}(\bm{\tilde{x}})\!-\!{\breve{\bm{\Delta}}}(\bm{\tilde{x}})\|&\leq\|\bm{\beta}\|\|\bm{\Sigma}_{\Delta}^{1/2}(\bm{\tilde{x}})\|\}\\
\quad\quad\quad\quad&\geq (1-\delta)^3
\end{aligned}
\ee
where $\bm{\mu}_{\Delta,k}(\bm{\tilde{x}})$ represents the  {predictive mean of the} recursively updated GP given by \eqref{34} {at time $k\in\mathbb{N}$}.
}
\end{proposition}

  % boundedness of estimation error

\YLnew{\LY{The proof of Proposition \ref{prop1} follows the lines of the proof of Lemma \ref{lemma:1}}.
Proposition \ref{prop1} ensures the boundedness of the GP regression error between the true function ${\breve{\bm{\Delta}}}(\bm{\tilde{x}})$ and the recursively updated predictive mean function $\bm{\mu}_{\Delta,k}(\bm{\tilde{x}})$ with a high probability, where the error bound is proportional to the predictive standard deviation \cite{umlauft2019feedback}.}

{Based on the online adapted GP, we propose the following controller:}
\begin{multline} %\label{eq:control}
\label{ctrl}
\bm{u}{[k]}= -\bm{\zeta}_\mathrm{p}(\breve{k}_\mathrm{p},\bm{\Sigma}_\Delta(\tilde{\bm{x}}{[k]}))\bm{q}_\mathrm{e}{[k]}-\bm{\zeta}_\mathrm{d}(\breve{k}_\mathrm{d},\bm{\Sigma}_\Delta(\tilde{\bm{x}}{[k]}))\bm{\omega}{[k]}\\
-\bm{J}_{\mathrm{c}0}\bm{\mu}_{\Delta,k}(\tilde{\bm{x}}{[k]})+\bm{\omega}^{\times}{[k]}\bm{J}_{\mathrm{c}0}\bm{\omega}{[k]}, 
%\bm{u}{(t)}= -\bm{\zeta}_\mathrm{p}(\breve{k}_\mathrm{p},\bm{\Sigma}_\Delta(\tilde{\bm{x}}{(t)}))\bm{q}_\mathrm{e}{(t)}-\bm{\zeta}_\mathrm{d}(\breve{k}_\mathrm{d},\bm{\Sigma}_\Delta(\tilde{\bm{x}}{(t)}))\bm{\omega}{(t)}\\
%-\bm{J}_{\mathrm{c}0}\bm{\mu}_{\Delta,k}(\tilde{\bm{x}}{(t)})+\bm{\omega}^{\times}{(t)}\bm{J}_{\mathrm{c}0}\bm{\omega}{(t)}, 
\end{multline}
where  %${t\in((k+1)T_\mathrm{s}, kT_\mathrm{s}]}$, 
${k\in\mathbb{N}}$ is the discrete time \TR{and} $[\bm{{\mu}}_{\Delta,k}]_j=\bm{\alpha}^{\top}_j[k] \bm{K}^{j}_{M+}$, ${j\in\mathbb{I}_1^3}$ %=1,2,3$ 
is the predictive mean updated at $k$-th {step} according to the ROSGP algorithm proposed in Section \ref{sec:3}.\ref{sec:3:b}.
%III.B, 
 {The functions $\bm{\zeta}_\mathrm{p}$ and $\bm{\zeta}_\mathrm{d}$ correspond to the feedback gains %functions in the controller to be 
of the proposed control law, parameterized in terms of $\breve{k}_\mathrm{p}\in\mathbb{R}^{n_\mathrm{p}}$, $\breve{k}_\mathrm{d}\in\mathbb{R}^{n_\mathrm{d}}$. These functions are chosen such that the following condition is satisfied:}

%{In this structure we assume that the attitude control is running on a much higher smapling rate than the update process for the GP, hence it can be considered as a continuous-time process compared to the GP updates happening at sampling moments $kT_\mathrm{s}$}

\begin{condition} \label{cond:bound}
\rm 
For given sets $\mathbb{S}_\ast\subseteq {\mathbb{S}}^{3\times 3}$ 
and $\mathbb{K}_\ast\subseteq \mathbb{R}$, the symmetric functions  
$\bm{\zeta}_\mathrm{p}, \bm{\zeta}_\mathrm{d}: \mathbb{S}^{3\times 3}\to\mathbb{S}^{3\times 3}$ 
are monotone increasing w.r.t. $K$ and bounded in the sense that the minimum 
$\underline{\zeta}_\tau$ of $\sigma_{\mathrm{min}}(\bm{\zeta}_{\tau}(K,\breve{k}_\tau)$
and the maximum  $\bar{\zeta}_\tau$ of $\sigma_{\mathrm{min}}(\bm{\zeta}_{\tau}(K,\breve{k}_\tau)$ 
exist over $K\in \mathbb{S}_\ast$ and $\breve{k}_\tau \in \mathbb{K}_\ast ^ {n_\tau}$ where 
$\sigma_{\mathrm{min}}$ and $\sigma_{\mathrm{max}}$ corresponds to the minimum and maximum singular values and 
$\tau \in\{\mathrm{p},\mathrm{d}\}$. Then
%symmetric %functions of trained GP variance $\bm{\Sigma}_\Delta$, 
%and 
%are 
% ^{3\times 3}
%bounded {in the sense that for any \LY{$K\in\bar{\mathbb{S}}$} and $\breve{k}_\mathrm{p}\in\mathbb{R}^{n_\mathrm{p}}$, $\breve{k}_\mathrm{d}\in\mathbb{R}^{n_\mathrm{d}}$, %\LY{$\lambda_{\mathrm{min}}(\bm{\zeta}_{\sigma}%{(\breve{k}_\mathrm{p},K)}
%)$ and $\lambda_{\mathrm{max}}(\bm{\zeta}_{\sigma}
%{(\breve{k}_\mathrm{p},K)}
%)$ are bounded where $\bar{\mathbb{S}}$ is a subset of ${\mathbb{S}}^{3\times 3}$, and $\bm{\zeta}_{\sigma}\in\{\bm{\zeta}_\mathrm{p}, \bm{\zeta}_\mathrm{d}\}$.} 
%
%
%\LY{That is,
%there exist $\underline{\zeta}_\mathrm{p},\bar{\zeta}_\mathrm{p},\underline{\zeta}_\mathrm{d},\bar{\zeta}_\mathrm{d} \in \mathbb{R}_0^+$ such that:}
\begin{align*}\underline{\zeta}_\mathrm{p}\|\bm{w}\|^2 &\leq\bm{w}^{\top}\bm{\zeta}_\mathrm{p}{(\breve{k}_\mathrm{p},K)}\bm{w}\leq \bar{\zeta}_\mathrm{p}\|\bm{w}\|^2\\
\underline{\zeta}_\mathrm{d}\|\bm{w}\|^2&\leq\bm{w}^{\top}\bm{\zeta}_\mathrm{d}{(\breve{k}_\mathrm{d},K)}\bm{w}\leq \bar{\zeta}_\mathrm{d}\|\bm{w}\|^2
\end{align*}
holds for all $\bm{w}\in\mathbb{R}^3$. 
\end{condition}
\vspace{-0.2cm}
\LY{There is a wide \TR{class} %selection
of  $\bm{\zeta}$ {functions such that Condition \ref{cond:bound} is satisfied. For instance, if ${\mathbb{S}_\ast}$ and $K_\ast$ are bounded sets, then polynomial functions can be selected for $\bm{\zeta}_{\sigma}$. Otherwise, they can be chosen as saturated functions such as sigmoid or Gauss functions.
}
Furthermore, $\breve{k}_\mathrm{p}$ and $\breve{k}_\mathrm{d}$ {can be seen as tuning parameters that can be adjusted based on} \LY{practical experience w.r.t. nominal controller design and do not require extensive tuning procedures.}}
%prior knowledge of the combined spacecraft dynamics (e.g., {$J_{\mathrm{c}0}$)} {or} reasonably selected based on the inertia of the servicer.

%\begin{remark} 
%{\rm \YLnew{The functions $\bm{\zeta}_{(\cdot)}$ {correspond to the} feedback gain{s} %functions in the controller to be 
%{of the} proposed {controller}, which are designed to be monotone increasing w.r.t. the predictive variance $\bm{\Sigma}_\Delta$, reflecting its relevance to feedback gain.
%There is a wide selection of  $\bm{\zeta}_{(\cdot)}$, such as the polynomial function, sigmoid function, and Gauss error function. Furthermore, $\breve{k}_\mathrm{p}$ and $\breve{k}_\mathrm{d}$ represent the prior knowledge of the combined spacecraft dynamics, i.e. they are reasonably selected based on the inertia of the servicer.}}
%\end{remark}

%\vspace{2mm}

%\begin{remark}
%{\rm \YLnew{It is worth mentioning that the system model \eqref{6} is given in a continuous differential equation, whereas the ROSGP is indeed updated in a discrete-time manner. Despite this, the discrete-time ROSGP algorithm is reasonable and implementable in practical engineering, as the working signals from the onboard computer are discrete in such environments. Alternatively, for small $T_{\mathrm{s}}$, \eqref{46} may be considered as a discrete approximation to a first-order differential equation in $\bm{\alpha}(t)$, and thus it is reasonable to assume that $\bm{\alpha}(t)=\bm{\alpha}[k]$ and $\bm{\alpha}(t-T_{\mathrm{s}})=\bm{\alpha}[k-1]$. 
%}}
%\end{remark}
\vspace{0.1cm}
The %specific 
implementation {of the controller} %procedure 
is summarized in Algorithm \ref{alg:1}.

\subsection{Stability Analysis}

{In order to conduct the stability analysis of the proposed scheme, we take the assumption that $T_\mathrm{s}$ is small enough such that $\bm{x}(t)\approx \bm{x}[k]$ for $t\in(kT_\mathrm{s},(k+1)T_\mathrm{s}]$, which is reasonable in case of $10$-$50$ Hz sampling based attitude control of a satellite. This simplifies our analysis as instead of sampled data \TR{related} issues we can focus on the interplay between the GP-based adaptive control law and the motion dynamics of the combined spacecraft. For this reason we will treat \eqref{ctrl} as a continuous time control law and consider $\bm{\alpha}$ varying continuously with time $t$.}

%{Take assumption of fast sampling time!}

% \YL{As shown in Algorithm \ref{alg:1}, the system runs in a continuous way while the GP model updates in a discrete manner, which leads to a switched continuous-time system. Thus, in this subsection, the convergence of the system states is firstly analyzed based on the principle of the common Lyapunov function\cite{liberzon2003switching}.}

%In this subsection, we analyze the stability of the closed-loop system employing the controller \eqref{ctrl}.

\begin{algorithm}[t]
  \caption{\YL{GP-based online learning control strategy for attitude takeover tasks}}
   \label{alg:1}
\begin{algorithmic}[1]
  \Statex \textbf{Initialization}: {Choose} $\bm{Q}_{0}$, $\bm{Q}_\mathrm{d}$, $\bm{\omega}_{0}$,~$\breve{k}_\mathrm{p}$,~$\breve{k}_\mathrm{d}$, $\lambda$, $\bm{P}[0]$, ~$k=0$, $T_\mathrm{s} $,  $N$. \vspace{0.2mm}
\Statex \textbf{Offline Training Phase}: \vspace{0.1mm}
\State Generate training data set $\mathcal{D}_N$;
\State Train GP model to {optimise} $\hat{\bm{\theta}}$ and $\bm{\alpha}[0]$.\vspace{1mm}
\Statex \textbf{Online Control Phase}:\vspace{0.1mm}
 \For{%each time moment 
$k=1,2,...,$}
    \State Observe the current $\tilde{\bm{x}}[k]$ {and $\bm{y}[k]$};
    %\State Compute the current ;
      \State Update:
       \State \quad $\Pk\! \gets  \!\lambda^{-1}\Pkfirst \!- \!\lambda^{-1}\!\Lk\!\bm{K}_{M{[k]}}^{\top}\Pkfirst$
      \State \quad $\Lk\gets\Pk\bm{K}_{M{[k]}}$
      \State \quad
      ${r}[k]\gets {y}[k]-\bm{\alpha}^{\top}[k-1]\bm{K}_{M{[k]}} $
      \State \quad $\alphak \gets \alphakfirst + \Lk {r}[k]$
      \State Compute $\bm{\mu}_{\Delta,k}$ and  $\bm{\Sigma}_{\Delta}$ with \eqref{34} and \eqref{32};
      \State Compute $\bm{u}{[k]}$ using \eqref{ctrl};
      \State \textls[-25]{Apply $\bm{u}(t)\!=\!\bm{u}{[k]}$ on the system for $t\!\in\![kT_\mathrm{s},\!(k+1)T_\mathrm{s}]$}
    \EndFor  
\end{algorithmic}
 \end{algorithm}

Assume that {the} {true} model uncertainty {$\breve{\bm{\Delta}}(\tilde{\bm{x}}(t))=\bm{J}_{\mathrm{c}0}^{-1}\bm{d}(\tilde{\bm{x}}(t))$ at time moment $t$} 
can be parameterized as $\bm{d}(\tilde{\bm{x}}{(t)})=\bm{J}_{\mathrm{c}0}\breve{\bm{\alpha}}^{\top}(t)\bm{K}_{M{(t)}}$, where $\breve{\bm{\alpha}}(t)$ denotes the true value of the weight for time $t$. Define $\tilde{\bm{\alpha}}(t)= \bm{\alpha}(t)-\breve{\bm{\alpha}}(t)$ as the deviation of the weight for the time moment $t$. Substituting the controller \eqref{ctrl} into \eqref{5} yields the closed-loop system:
\vspace{-1mm}
\begin{multline}
\label{52}
\bm{J}_{\mathrm{c}0}\dot{\bm{\omega}}{(t)} = -\bm{\zeta}_\mathrm{p}(\breve{k}_\mathrm{p},\bm{\Sigma}_\Delta{(t)})\bm{q}_\mathrm{e}{(t)}-\bm{\zeta}_\mathrm{d}(\breve{k}_\mathrm{d},\bm{\Sigma}_\Delta{(t)})\bm{\omega}{(t)}\\
-\bm{J}_{\mathrm{c}0}\tilde{\bm{\alpha}}^{\top}\!(t)\bm{K}_{M{(t)}}.
\end{multline}

%Then the 
\TR{The} following theorem %is given to show 
\TR{shows} %the 
boundedness of the closed-loop signals.

\begin{theorem} \label{th1}
{\rm Consider that the combined spacecraft \eqref{6} {which} satisfies {Conditions \ref{cond:1}-\ref{assump:4} and Assumption \ref{assmp:1}}. Under the proposed GP-based learning control {law} \eqref{ctrl}, {satisfying Condition \ref{cond:bound}}, where the unknown function $\breve{\bm{\Delta}}$ is modeled by a multi-dimensional GP \eqref{11} {that} is recursively online updated according to Algorithm \ref{alg:1}, $\bm{Q}_\mathrm{e}$, $\bm{\omega}$  and $\tilde{\bm{\alpha}}$ are guaranteed to be ultimately uniformly bounded.}% and converg{ing} to a set $\Omega_0$ {around} the origin.  }
\end{theorem}

\begin{proof}
Consider the radially unbounded Lyapunov function:
\vspace{-2mm}
\be
\label{53}
V_{0}(t) = \frac{1}{2}\bm{\omega}^{\top}{(t)}\bm{J}_{\mathrm{c}0}\bm{\omega}{(t)} 
+\int_{t-T_{\mathrm{s}}}^{t} \!\!\tilde{\bm{\alpha}}^{\top}({\tau})\tilde{\bm{\alpha}}({\tau})\mathrm{d}{\tau}.
\ee
{By} differentiating \eqref{53} along the system trajectories \eqref{52}, 
\begin{align}
\dot{V}_{0}(t) &=\!\bm{\omega}^{\top}\!{(t)}(-\bm{\zeta}_\mathrm{p}(\breve{k}_\mathrm{p},\bm{\Sigma}_\Delta{(t)})\bm{q}_\mathrm{e}{(t)}\!-\!\bm{\zeta}_\mathrm{d}(\breve{k}_\mathrm{d},\!\bm{\Sigma}_\Delta{(t)})\bm{\omega}{(t)}
\notag\\ &~~-\!\bm{J}_{\mathrm{c}0}\tilde{\bm{\alpha}}^{\!\top}\!(t)\bm{K}_{M{(t)}})\!
+\!\tilde{\bm{\alpha}}^{\!\top}\!(t)\tilde{\bm{\alpha}}(t)-
\tilde{\bm{\alpha}}^{\!\top}\!(t-T_{\mathrm{s}})\tilde{\bm{\alpha}}(t-T_{\mathrm{s}}) \notag\\
&\leq -\underline{\zeta}_\mathrm{d}\|\bm{\omega}{(t)}\|^2-\bm{\omega}^{\top}\!\bm{\zeta}_\mathrm{p}(\breve{k}_\mathrm{p},\bm{\Sigma}_\Delta{(t)})\bm{q}_\mathrm{e}{(t)} \notag\\ 
&~~-\bm{\omega}^{\!\top}\!{(t)}\bm{J}_{\mathrm{c}0}\tilde{\bm{\alpha}}^{\!\top}\!(t)\bm{K}_{M{(t)}}
+\tilde{\bm{\alpha}}^{\!\top}\!(t)\tilde{\bm{\alpha}}(t)\notag \\ &~~-
\tilde{\bm{\alpha}}^{\!\top}\!(t-T_{\mathrm{s}})\tilde{\bm{\alpha}}(t-T_{\mathrm{s}}). \label{54}
\end{align}
Furthermore, from \eqref{46}, we have:
\begin{align}
\tilde{\bm{\alpha}}(t)&=\bm{\alpha}(t-T_{\mathrm{s}}) + \bm{L}(t)r(t)-\breve{\bm{\alpha}}(t) \label{53-1}, \\
&=\tilde{\bm{\alpha}}(t-T_{\mathrm{s}}) + \bm{L}(t)r(t)+\breve{\bm{\alpha}}(t-T_{\mathrm{s}})-\breve{\bm{\alpha}}(t). \notag
\end{align}
Define $\bm{\Psi}{(t)}=\breve{\bm{\alpha}}(t-T_{\mathrm{s}})-\breve{\bm{\alpha}}(t)$ and $\bm{\Theta}{(t)}=\bm{L}(t)r(t)$. Then, it follows that:
\begin{multline}
\tilde{\bm{\alpha}}^{\!\top}\!(t)\tilde{\bm{\alpha}}(t)=\tilde{\bm{\alpha}}^{\!\top}\!(t-T_{\mathrm{s}})\tilde{\bm{\alpha}}(t-T_{\mathrm{s}})
+ 2\tilde{\bm{\alpha}}^{\top}(t-T_{\mathrm{s}})\bm{\Theta}{(t)}\\
+2\tilde{\bm{\alpha}}^{\!\top}\!(t-T_{\mathrm{s}})\bm{\Psi}{(t)} + 2\bm{\Theta}^{\!\top}\!{(t)}\bm{\Psi}{(t)}. \label{53-2}
\end{multline}
Using the special case of Young's inequality $\bm{x}^{\top}{\bm{y}}\leq \frac{1}{2}\bm{x}^{\top}\bm{x} +\frac{1}{2}\bm{y}^{\top}\bm{y}$, the following inequalities can be derived:
\begin{subequations}
\begin{align}
\label{53-3}
2\tilde{\bm{\alpha}}^{\!\top}\!(t-T_{\mathrm{s}})\bm{\Theta}{{(t)}}&\leq \|\tilde{\bm{\alpha}}(t-T_{\mathrm{s}})\|^2 + \|\bm{\Theta}{{(t)}}\|^2 \\
%\tilde{\bm{\alpha}}^{\top}(t-T_{\mathrm{s}})\tilde{\bm{\alpha}}(t-T_{\mathrm{s}}) + \bm{\Theta}^{\!\top}\!{{(t)}}\bm{\Theta}{{(t)}}\\
\label{53-3-1}
2\tilde{\bm{\alpha}}^{\!\top}\!(t-T_{\mathrm{s}})\bm{\Psi}{{(t)}}&\leq \|\tilde{\bm{\alpha}}(t-T_{\mathrm{s}})\|^2 + \|\bm{\Psi}{{(t)}}\|^2 \\
%\tilde{\bm{\alpha}}^{\top}(t-T_{\mathrm{s}})\tilde{\bm{\alpha}}(t-T_{\mathrm{s}})+\bm{\Psi}^{\!\top}\!{{(t)}}\bm{\Psi}{{(t)}}\\
\label{53-3-2}
-\bm{\omega}^{\!\top}\!{{(t)}}\bm{\zeta}_\mathrm{p}(\breve{k}_\mathrm{p},\bm{\Sigma}_\Delta{{(t)}})\bm{q}_\mathrm{e}{{(t)}}&\leq\frac{1}{2}\bar{{\zeta}}_\mathrm{p}^2+\frac{1}{2}\|\bm{\omega}{{(t)}}\|^2\\
\label{53-3-3}
-\bm{\omega}^{\!\top}\!{{(t)}}\bm{J}_{\mathrm{c}0}\tilde{\bm{\alpha}}^{\!\top}\!(t)\bm{K}_{M{(t)}}&\leq\frac{1}{2}\lambda_J^2\bar{\sigma}_\mathrm{f}^4\|\tilde{\bm{\alpha}}(t)\|^2+\frac{1}{2}\|\bm{\omega}{{(t)}}\|^2
\end{align}
\end{subequations}
where $\bar{\sigma}_\mathrm{f}=\mathrm{sup}~\sigma_{\mathrm{f},j}$, $j\in\mathbb{I}_1^3$.
Substituting the inequalities \eqref{53-3}-\eqref{53-3-3} into \eqref{54} leads to:
\begin{multline}
\label{53-4}
\dot{V}_{0}(t) \leq-(\underline{\zeta}_\mathrm{d}-1)\|\bm{\omega}{{(t)}}\|^2-\gamma\|\tilde{\bm{\alpha}}(t)\|^2+\frac{1}{2}\bar{\zeta}_\mathrm{p}^2\\
+(\frac{1}{2} \lambda_J^2 \bar{\sigma}_{\mathrm{f}}^4+1+\gamma) (3 \|\tilde{\bm{\alpha}}(t-T_{\mathrm{s}})\|^2+3\|\bm{\Psi}(t)\|^2+3\|\bm{\Theta}(t)\|^2 )\\
-\|\tilde{\bm{\alpha}}(t-T_{\mathrm{s}})\|^2,
\end{multline}
where $\gamma$ is a positive constant. Furthermore, according to the definition of $\bm{\Psi}$ and $\bm{\Theta}$, it is reasonable to assume that $\bm{\Psi}$ and $\bm{\Theta}$ are bounded\footnote{\LY{Generally, the unknown uncertainties will not vary with an infinite rate for on-orbit scenarios, thus $\|\bm{\Psi}(t)\|$ has an upper bound. Moreover, according to the definition of $\bm{\Phi}$, if the condition of persistent excitation is satisfied (which is mild for attitude dynamics), then $\|\bm{\Phi}(t)\|$ is also bounded.}}, that is, $\|\bm{\Psi}{(t)}\|\leq\bar{\phi}$ and $\|\bm{\Theta}{(t)}\|\leq\bar{\xi}$. If $\underline{\zeta}_\mathrm{d}-1>0$ and $1-3(\frac{1}{2} \lambda_J^2 \bar{\sigma}_{\mathrm{f}}^4+1+\gamma)>0$ are satisfied, then the time derivative of $V_0$ can be given as:
\be
\label{53-5}
\dot{V}_{0}(t) \leq -\eta\|\bm{\omega}{(t)}\|^2-\gamma\|\tilde{\bm{\alpha}}(t)\|^2+\varpi
\ee
where $\eta=\underline{\zeta}_\mathrm{d}-1$, $\varpi=\frac{1}{2} \bar{\zeta}_\mathrm{p}^2+3(\frac{1}{2} \lambda_J^2 \bar{\sigma}_{\mathrm{f}}^4+1+\gamma)(\bar{\phi}^2+\bar{\xi}^2)$. Clearly,  equation \eqref{53-5} indicates that $\bm{\omega}{(t)}$ and $\tilde{\bm{\alpha}}(t)$ are ultimately uniformly bounded {in the set}:
\be
\label{56}
\Omega_0\!=\!\!\left\{ {(\bm{\omega},\tilde{\bm{\alpha}}) \in \mathbb{R}^{3} \times \mathbb{R}^3} \Big|
%\forall\bm{\omega}\in\mathbb{R}^3,\tilde{\bm{\alpha}}(t)\in\mathbb{R}^M\Big|~\
\|\bm{\omega}\|^{2}\leq\frac{\varpi}{\eta}, \|\tilde{\bm{\alpha}}\|^{2}\leq\frac{\varpi}{\gamma}\right\}
\ee

Furthermore, in view of the natural boundedness of the quaternion, one can conclude that the system state $(\bm{Q}_\mathrm{e},\bm{\omega})$ is ultimately uniformly bounded.
\end{proof}
\vskip 2mm

\YLnew{Theorem \ref{th1} implies the existence of the ultimate bound of system states $(\bm{Q}_\mathrm{e},\bm{\omega})$. However, because $\varpi$ is an unknown constant, the specific size of the ultimate bound is hard to know. Next, we will make a step further to quantify the ultimate bound of system states $(\bm{Q}_\mathrm{e},\bm{\omega})$ by taking Proposition \ref{prop1} into consideration.}

\begin{theorem} 
{\rm  Consider that the combined spacecraft \eqref{6} {which} satisfies {Conditions \ref{cond:1}-\ref{assump:4} and Assumption \ref{assmp:1}}. Under the proposed GP-based learning control {law} \eqref{ctrl}, {satisfying Condition \ref{cond:bound} and}
%Consider that the combined spacecraft system \eqref{6} satisfies Assumptions 1-5. Under the proposed GP-based learning controller \eqref{ctrl}, 
where the unknown function $\breve{\bm{\Delta}}$ is modeled by a multi-dimensional GP \eqref{11} {that} is recursively online updated according to Algorithm \ref{alg:1}, %Then 
$\bm{Q}_\mathrm{e}$ and $\bm{\omega}$ are guaranteed to be ultimately uniformly bounded %and {converging} to the small sets $\Omega_1$ and $\Omega_2$ respectively
with a probability of $(1-\delta)^3$.}
\end{theorem} 

\begin{proof} 
Consider the Lyapunov {function} candidate as:
\be
\label{57}
\begin{aligned}
V_{1} &= (\bm{\zeta}_\mathrm{p}(\breve{k}_\mathrm{p},\bm{\Sigma}_\Delta)+\nu\bm{\zeta}_\mathrm{d}(\breve{k}_\mathrm{d},\bm{\Sigma}_\Delta))((1-q_{\mathrm{e}0})^{2}+\bm{q}_\mathrm{e}^{\top}\bm{q}_\mathrm{e})\\
&~~+\frac{1}{2}\bm{\omega}^{\top}\bm{J}_{\mathrm{c}0}\bm{\omega}+\nu\bm{q}_\mathrm{e}^{\top}\bm{J}_{\mathrm{c}0}\bm{\omega},  
\end{aligned}
\ee
where  $\nu>0$ {is a constant}. The proof for the positive-definiteness of $V_1$ can be found in \cite{gui2018robustness}.

% \be
% \label{23}
% \begin{matrix}
% V \geq (\bm{K}_\mathrm{p}+\nu\bm{K}_\mathrm{d})\bm{q}^{\top}\bm{q}+\frac{1}{2}\bm{\omega}^{\top}\bm{J}\bm{\omega}+\nu\bm{q}^{\top}\bm{J}\bm{\omega}\\
% =\bm{x}^{\top}
% \end{matrix}
% \ee

By differentiating $V_{1}$ and employing the closed-loop system and controller, one can derive:
\be
\label{58}
\begin{aligned}
\dot{V}_{1} &= 
% (\bm{\zeta}_\mathrm{p}(\breve{k}_\mathrm{p},\bm{\Sigma}_\Delta)+\nu\bm{\zeta}_\mathrm{d}(\breve{k}_\mathrm{d},\bm{\Sigma}_\Delta))\bm{q}_\mathrm{e}^{\top}\bm{\omega}+\bm{\omega}^{\top}\bm{J}_{\mathrm{c}0}\dot{\bm{\omega}}\\
% &\quad +\nu\dot{\bm{q}}_\mathrm{e}^{\top}\bm{J}_{\mathrm{c}0}\bm{\omega}+\nu\bm{q}_\mathrm{e}^{\top}\bm{J}_{\mathrm{c}0}\dot{\bm{\omega}}\\
% &=
(\bm{\zeta}_\mathrm{p}(\breve{k}_\mathrm{p},\bm{\Sigma}_\Delta)+\nu\bm{\zeta}_\mathrm{d}(\breve{k}_\mathrm{d},\bm{\Sigma}_\Delta))\bm{q}_\mathrm{e}^{\top}\bm{\omega}\\
&\quad +(\bm{\omega}+\nu\bm{q}_\mathrm{e})^{\top}(-\bm{\zeta}_\mathrm{p}(\breve{k}_\mathrm{p},\bm{\Sigma}_\Delta)\bm{q}_\mathrm{e}-\bm{\zeta}_\mathrm{d}(\breve{k}_\mathrm{d},\bm{\Sigma}_\Delta)\bm{\omega}\\
&\quad-\bm{J}_{\mathrm{c}0}\bm{\mu}_{\Delta,k}(\tilde{\bm{x}})+\bm{d}(\tilde{\bm{x}}))+\frac{1}{2}\nu\bm{\omega}^{\top}(q_{\mathrm{e}0}\bm{I}_{3}+\bm{q}_\mathrm{e}^{\times})\bm{J}_{\mathrm{c}0}\bm{\omega}
\end{aligned} 
\ee
According to Cauchy-Schwartz inequality and $\|q_{\mathrm{e}0}\bm{I}_{3}+\bm{q}_\mathrm{e}^{\times}\|=1$:
\be
\label{59}
\begin{aligned}
\dot{V}_{1} &= -\nu\bm{\zeta}_\mathrm{p}(\breve{k}_\mathrm{p},\bm{\Sigma}_\Delta)\|\bm{q}_\mathrm{e}\|^{2}-\bm{\zeta}_\mathrm{d}(\breve{k}_\mathrm{d},\bm{\Sigma}_\Delta)\|\bm{\omega}\|^{2}\\
&~~~+\frac{1}{2}\nu\bm{\omega}^{\top}(q_{\mathrm{e}0}\bm{I}_{3}+\bm{q}_\mathrm{e}^{\times})\bm{J}_{\mathrm{c}0}\bm{\omega}\\
&\quad+(\bm{\omega}+\nu\bm{q}_\mathrm{e})^{\top}(\bm{d}(\tilde{\bm{x}})-\bm{J}_{\mathrm{c}0}\bm{\mu}_{\Delta,k}(\tilde{\bm{x}}))\\
& \leq -\nu \underline{\zeta}_\mathrm{p}\|\bm{q}_\mathrm{e}\|^{2}-\lambda_{J}(\underline{\zeta}_\mathrm{d}-\frac{1}{2}\nu)\|\bm{\omega}\|^{2}\\
&~~~+(\bm{\omega}+\nu\bm{q}_\mathrm{e})^{\top}(\bm{d}(\tilde{\bm{x}})-\bm{J}_{\mathrm{c}0}\bm{\mu}_{\Delta,k}(\tilde{\bm{x}}))
\end{aligned} 
\ee

Further combining Proposition \ref{prop1} with \eqref{59}, it can be deduced that:
\be
\label{60}
\begin{aligned}
\mathbb{P}\{\dot{V}_{1}&\leq-\lambda_{\rm{min}}(\bm{M}_{l})\|\bm{x}\|^{2}+\lambda_{J}(\|\bm{\omega}\|+\nu\|\bm{q}_\mathrm{e}\|)\|\bm{\beta}\|\|\bm{\Sigma}^{1/2}_{\Delta}\| \}\\
&\qquad\qquad\qquad\geq(1-\delta)^3
\end{aligned}
\ee
{holds for all $t\in\mathbb{R}_0^+$} where 
\be
\label{61}
\bm{M}_{l}=\left[\begin{array}{cc}
           \nu \underline{\zeta}_\mathrm{p} & 0\\
                0     &\underline{\zeta}_\mathrm{d}-\frac{1}{2}\nu\lambda_{J}
       \end{array}\right].
\ee

From the definition of ${V}_{1}$, we have:
\be
\label{62}
\begin{aligned}
{V}_{1}\leq \bm{x}^{\top}\bm{M}_{s}\bm{x}\leq\lambda_{\rm{max}}(\bm{M}_{s})\|\bm{x}\|^{2}, \\
\quad \bm{M}_{s} = \left[\begin{array}{cc}
2(\bar{\zeta}_\mathrm{p}+\nu \bar{\zeta}_\mathrm{d}) & \frac{1}{2}\nu\lambda_{J}\\
\frac{1}{2}\nu\lambda_{J} & \frac{1}{2}\lambda_{J}
\end{array}\right]
\end{aligned}
\ee
Moreover, we can derive the boundedness of $\bm{q}_\mathrm{e}$ and $\bm{\omega}$, which indicates that:
\be
\label{63}
\|\bm{q}_\mathrm{e}\|\leq\sqrt{\frac{{V}_{1}}{\bar{\zeta}_\mathrm{p}+\nu \bar{\zeta}_\mathrm{d}}}, \quad 
\|\bm{\omega}\|\leq\sqrt{\frac{2{V}_{1}}{\lambda_{c}}}.
\ee

Consequently, from \eqref{60}, \eqref{62} and \eqref{63}, one can readily derive:
\begin{multline}
\label{64}
\mathbb{P}  \bigg\{\dot{V}_{1}\leq-\frac{\lambda_{\rm{min}}(\bm{M}_{l})}{\lambda_{\rm{max}}(\bm{M}_{s})}{V}_{1}\\+\lambda_{J}\left(\sqrt{\frac{2{V}_{1}}{\lambda_{c}}}+\nu\sqrt{\frac{{V}_{1}}{\bar{\zeta}_\mathrm{p}+\nu \bar{\zeta}_\mathrm{d}}}\right)\varepsilon
\bigg\}
\geq (1-\delta)^3,
\end{multline}
{holds for all $t\in\mathbb{R}_0^+$}  where $\varepsilon\ge  \|\bm{\beta}\|\|\bm{\Sigma}_\Delta^{1/2}\|$. 

\vspace{2mm}
Next, we will take a step further to estimate the ultimately uniform bound for attitude error $\bm{q}_\mathrm{e}$ and angular velocity $\bm{\omega}$. Defining $W=\sqrt{V_{1}}$, we have \TR{that} $\dot{W}=\dot{{V}}_{1}/(2\sqrt{{V}_{1}})$ holds when $V_{1}\not=0$, that is:
\be
\label{65}
\mathbb{P}\left\{\dot{W}\leq-\frac{\lambda_{\rm{min}}(\bm{M}_{l})}{2\lambda_{\rm{max}}(\bm{M}_{s})}{W}+\frac{\lambda_{J}\varepsilon}{\sqrt{2}}\vartheta
\right\}\geq(1-\delta)^3
\ee
with $\vartheta = \left(\sqrt{\frac{1}{2\lambda_{c}}}+\nu\sqrt{\frac{1}{2(\bar{\zeta}_\mathrm{p}+\nu \bar{\zeta}_\mathrm{d})}}\right)$. Integrating the inequality in \eqref{65} leads to:
\be
\label{66}
0\leq \sqrt{V_{1}(t)} \leq \left(\sqrt{V_{1}(0)}
-s_1
\right)e^{-s_{2}t}+ s_1
\ee
where 
%\begin{subequations}
%\begin{align}
\[
s_{1}=\frac{\sqrt{2}\lambda_{J}\varepsilon\vartheta\lambda_{\rm{max}}(\bm{M}_{s})}{\lambda_{\rm{min}}(\bm{M}_{l})}, \quad
s_{2}=\frac{\lambda_{\rm{min}}(\bm{M}_{l})}{2\lambda_{\rm{max}}(\bm{M}_{s})}.
\]
%\end{align}
%\end{subequations}
Thus, when $t\to\infty$, one can derive:
\be
\label{67}
\lim_{t\to\infty}\sup\sqrt{V_{1}(t)}\leq s_{1}
\ee
Recalling \eqref{63}, we can finally compute the bound of $\bm{q}_\mathrm{e}$ and $\bm{\omega}$ that:
\begin{subequations}
\begin{align}
\label{68}
&\lim_{t\to\infty}\sup \|\bm{q}_\mathrm{e}{(t)}\|=\frac{s_1}{\sqrt{\bar{\zeta}_\mathrm{p}+\nu \bar{\zeta}_\mathrm{d}}},\\
&\lim_{t\to\infty}\sup\|\bm{\omega}{(t)}\|=\sqrt{\frac{2}{\lambda_{c}}}s_1.
\end{align}
\end{subequations}
Furthermore, {due to} $2(\bar{\zeta}_\mathrm{p}+\nu \bar{\zeta}_\mathrm{d})(1-q_{\mathrm{e}0})\leq V_{1}$, we have:
\begin{align}
 \lim_{t\to\infty}\sup \|q_{\mathrm{e}0}{(t)}\| &\geq 1-\frac{\lim_{t\to\infty}\sup{V_{1}(t)}}{2(\bar{\zeta}_\mathrm{p}+\nu \bar{\zeta}_\mathrm{d})}\notag \\
 &\geq 1-\frac{s_1^{2}}{2(\bar{\zeta}_\mathrm{p}+\nu \bar{\zeta}_\mathrm{d})} .\label{69}
 \end{align}
Then, 
\begin{align}
\lim_{t\to\infty}\sup \|\bm{q}_\mathrm{e}{(t)}\|&=\lim_{t\to\infty}\sup \sqrt{1-q_{\mathrm{e}0}^{2}}\notag \\
&\leq \sqrt{1-(1-\frac{s_1^{2}}{2(\bar{\zeta}_\mathrm{p}+\nu \bar{\zeta}_\mathrm{d})})^{2}}. \label{70}
\end{align}

Comparing \eqref{70} and \eqref{68}, it can be concluded that the upper bound for $\bm{q}_\mathrm{e}$ is further reduced. Hence, the system states $\bm{q}_\mathrm{e}$ and $\bm{\omega}$ ultimately converge to the compact set $\Omega_1$ and $\Omega_2$ within the probability $(1-\delta)^3$:
%\label{71}
\begin{align*}
\Omega_1&=\left\{\bm{Q}_\mathrm{e}\in{\mathbb{Q}}^{3}\ \Big|\ \|\bm{q}_\mathrm{e}\|\leq\sqrt{1-(1-\frac{s_1^{2}}{2(\bar{\zeta}_\mathrm{p}+\nu \bar{\zeta}_\mathrm{d})})^{2}}\right\}\\
\Omega_2&=\left\{\bm{\omega}\in\mathbb{R}^3\ \Big|\ \|\bm{\omega}\|\leq\sqrt{\frac{2}{\lambda_{c}}}s_1 \right\}
\end{align*}
This completes the proof.
\end{proof}

\begin{figure*}[htpb]
\centering\includegraphics[width= 0.75\textwidth]{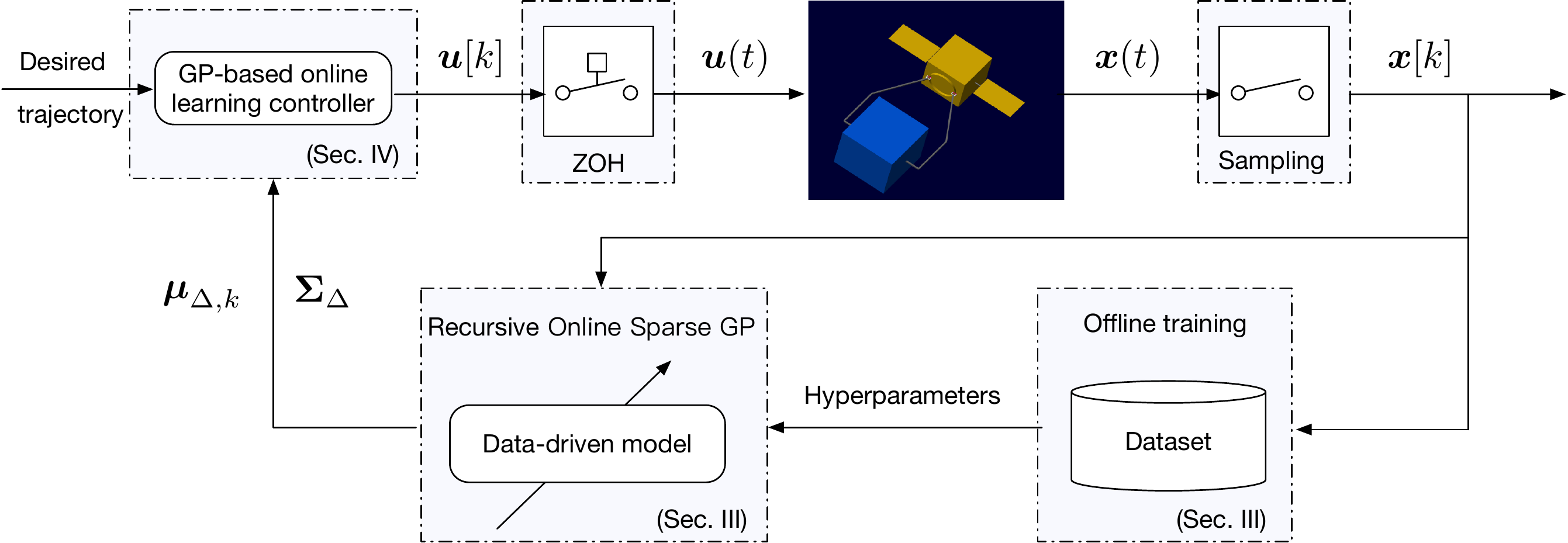} \vspace{2mm}
\caption{Block diagram of the proposed control strategy.}
\label{fig:3}
\end{figure*}

The block diagram of the closed-loop control system is shown in Fig.~\ref{fig:3}.

\begin{remark}
{\rm \RV{{Compared to} the existing attitude takeover controllers for the combined spacecraft, such as NN-based compensation \cite{huang2016impact} and adaptive control in \cite{guohua2020adaptive}, the advantages of the proposed GP-based online learning controller are as follows: 

1) As a nonparametric modeling approach, the predictive outputs of GP are inherently probabilistic. %, i.e. both the predictive mean and variance can be embedded into the controller design. 
%Wherein,
\RV{%In this context, 
\TR{The} GP variance quantifies the confidence level of the \TR{predicted uncertainty}, \TR{while the mean corresponds to the estimated \YLRV{uncertainty. Both the mean and the variance are incorporated into the control scheme, where the latter}}} effectively improves the robustness of the algorithm. To the best of the authors' knowledge, existing results for the uncertainty quantification of NN mainly include Monte Carlo method\YLRV{s}, dropout, and Bayesian ANNs. These \YLRV{approaches are generally complex} and computationally {too} demanding {to be applied in a real implementation on a spacecraft.}

2) In contrast \YLRV{to the manual time-consuming \TR{tuning} process for the hyperparameters of} the NN-based adaptive control scheme, the hyperparameters \YLRV{of the GP scheme, such as $\bm{\theta}_j=\operatorname{vec}(\sigma_{\epsilon, j}^2, \sigma_{\mathrm{f}, j}^2, \lambda_{j, 1}, \ldots, \lambda_{j, 9})$, are automatically tuned by marginalized likelihood based optimization, while the remaining parameters of the adaptive feedback controller, such as $\breve{k}_\mathrm{p}$ and $\breve{k}_\mathrm{d}$, can be designed by practical experience with PD control design for such systems with known nominal inertia matrix $\bm{J}_{\mathrm{c0}}$ and does not require extensive tuning procedures.}

3) As a data-driven \YLRV{adaptive approach}, the proposed GP-based learning controller \YLRV{only needs} an initial ``rough" model of the unknown dynamics \YLRV{that is learnt with only a small amount of data and {needs no extensive exploration of} the entire state space before {safe execution of} the mission. Subsequently, the initial GP model is updated online as new operational data becomes available to ensure continuous adaption and performance improvement of the controller. }
}}
\end{remark}

\section{Simulation Results}
\label{sec:sim}
In this section, simulation results under {various} on-orbit scenarios are presented to illustrate the effectiveness of our proposed control strategy. 

\subsection{Simulation Platform} % Introduction}
\begin{figure*}[htbp]
%\centering
%\begin{minipage}{0.45\textwidth}
\centering
\includegraphics[scale=0.65]{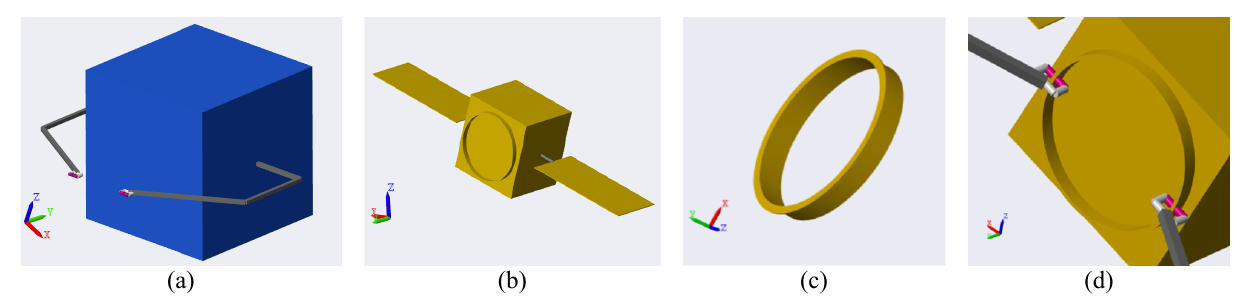}  
\caption[config]{\YLRV{Parts the combined spacecraft assembly: (a) Servicer spacecraft, (b) Target spacecraft, (c) Docking ring on the target, (d)  Capture point.}}
\label{config}
%\end{minipage}
\end{figure*}

\YLnew{A SimMechanics-based, high-fidelity simulator has been developed for the combined spacecraft to accurately characterize its attitude motion under the considered model uncertainties and target maneuvering, and act as the {physical system} and \emph{data-generator} for the proposed online learning control strategy. As shown in Fig.~\ref{config}, the system consists of two components, namely the servicer and the target.
The servicer, simplified as a cubesat, is equipped with two robotic arms on each side for capturing the target. Each arm is composed of three segments, and once capture is completed, the joints are locked in a specific configuration. The target is also modeled as a cube with solar panels on both sides and a capture interface in the form of a docking ring.
} \RV{For the simplicity of the study, each component in the system is considered as a rigid body without flexibility.}

{The} simulation platform can be seen as a black-box system, i.e. the attitude motion equations in the analytical form are ``packaged" within Simulink, and only the I/O ports are used for the proposed adaptive control algorithm, just as it would be the case for a real spacecraft. \YLRV{The physical parameters considered in the simulator are given in Table.~\ref{tab:1}, which are mainly \TR{based on} %referred to 
\cite{huang2016reconfigurable}. Note that appropriate modifications {have been done} to make {this scenario} more suitable for {implementation in} Matlab/SimMechanics.}

\begin{table}[]
\centering
\caption{Physical parameters of the combined spacecraft}
\small
\resizebox{\linewidth}{!}{
\begin{tabular}{llll}
\hline
\specialrule{0em}{1.0pt}{1.0pt}
\hline
                          &                                                                         & Parameter                        & Value                         \\ \hline
\multirow{6}{*}{Servicer} & \multirow{3}{*}{Body}                                                   & Size (m)                         & $2\times2\times2$             \\
                          &                                                                         & Inertia ($\rm{kg}\cdot m^2$) & $\rm{diag}(405, 405, 405)$    \\
                          &                                                                         & Mass (kg)                        & 1080                          \\ \cline{2-4} 
                          & \multirow{3}{*}{\begin{tabular}[c]{@{}l@{}}Robotic \\ Arm\end{tabular}} & Length per link (m)                         & 0.73, 1, 2                    \\
                          &                                                                         & Mass per link (kg)                         & 6                             \\
                          &                                                                         & Inertia per link ($\rm{kg}\cdot m^2$)                   & $\rm{diag}(0.03, 0.73, 0.73)$ \\ \hline
\multirow{5}{*}{Target}   & \multirow{3}{*}{Body}                                                   & Size (m)                         & $2\times2\times2$             \\
                          &                                                                         & Inertia ($\rm{kg}\cdot m^2$) & $\rm{diag}(36.8, 37.5, 36.8)$ \\
                          &                                                                         & Mass (kg)                        & 75                            \\ \cline{2-4} 
                          & \multirow{2}{*}{Others}                                                 & Size of solar panels (m)             & $1.5\times0.8\times0.01$      \\
                          &                                                                         & Size of docking ring (m)                     & 0.423/0.443 (I/O)                    \\ \hline
\end{tabular}}
\label{tab:1}
\end{table}

\YLRV{The studied simulation scenario is as follows. }The target is a partially malfunctioning satellite working in a sun-oriented mode. Thus, after being captured by the servicer, the target attempts to perform active attitude maneuvers throughout the takeover task, i.e., when the servicer applies a control torque to the target that leads to a deviation from its initial attitude orientation, the target will generate a competitive torque against the control torque to keep its attitude. The unknown model uncertainties considered in this example consist of two parts, \YLRV{the unknown dynamics \TR{corresponding to additional} %in terms of the 
robot arms, \TR{flexible} solar panels, etc.,}
%in addition to the nominal rigid body model,
and the additional dynamics caused by the target-generated torque. The active attitude control law for the target  is chosen to be of the PD form:\vspace{-1mm}
\be
\label{eq:target:cont}
\bm{u}_\mathrm{t} = -\bm{K}_\mathrm{pt}\bm{q}_\mathrm{et}-\bm{K}_\mathrm{dt}\bm{\omega}_\mathrm{t}
\ee
%\vspace{-1mm}
{where the subscript ``t" denotes the target-related variables}, $\bm{K}_\mathrm{pt} = 0.02\bm{J}_\mathrm{t}$, $\bm{K}_\mathrm{dt} = 0.05\bm{J}_\mathrm{t}$, $\bm{q}_\mathrm{et} = \bm{q}_\mathrm{dt}^{-1}\otimes\bm{q}_\mathrm{t}$, \YLRV{$\bm{q}_\mathrm{dt}$ is the vector part of initial quaternion of the target relative to the inertial coordinate system $\mathcal{F}_\mathrm{I}$.}
%representation in $\mathcal{F}_\mathrm{Bt}$ of the initial attitude in the body frame $\mathcal{F}_\mathrm{Bt}$ of the target.
%quaternion vector part of the target body coordinate system $\mathcal{F}_\mathrm{Bt}$ relative to the inertial coordinate system $\mathcal{F}_\mathrm{I}$. 
{Note that \eqref{eq:target:cont} is part of the black-box simulation system and it is not known by the servicer.}
% According to the configuration in SimMechanics, we know that $\bm{q}_{td}=[-0.251,~ 0.0235,~0.084]^{\top}$.

\subsection{Adaptive Control Under Unknown Model Uncertainties}

After docking, the initial Euler angle for the combined spacecraft is $[15 ~5 ~ -20]^{\circ}$, and the initial angular velocity is  $\bm{\omega}_{0}=[0.01~ 0.02~ -0.01]^{\top}$ rad/s. In this simulation scenario, the attitude orientation of the combined spacecraft is forced to converge to $\bm{Q}_\mathrm{d}=[1~ 0~ 0~ 0]^{\top}$, which corresponds to an attitude stabilization task.
\LY{Generally, there are two options to construct the training data set $\mathcal{D}_N$: Either just using the transient data of the closed-loop system or applying excitation torque on the spacecraft.  In this paper, only the transient data regarding orientation change with a baseline PD controller is collected, which represents a low-profile scenario. That is, the training data is collected during the control process in the first 50s of the simulation. }
%an excitation torque {under a} PD {controller} is applied to the combined spacecraft in the first 50s of the simulation, 
\YLRV{With a sampling frequency of 10 Hz, the size of the collected training set is $N = 500$. } %It is worth mentioning that the excitation torque is also used as the baseline control law for subsequent comparative simulations. 
Considering the sensor-based measurement of $\bm{q}_\mathrm{e}$ and $\bm{\omega}$, \YLRV{the training outputs are corrupted by {a} Gaussian white noise $\epsilon[k]$ with $\epsilon[k]\sim\mathcal{N}(0,0.05)$. }The feedback gain for the baseline controller follows $\bm{K}_\mathrm{p0}=0.1\bm{J}_{\mathrm{c}0}$, $\bm{K}_\mathrm{d0}=0.3\bm{J}_{\mathrm{c}0}$, where the nominal inertia matrix of the combined spacecraft is selected as $\bm{J}_{\mathrm{c}0}=\rm{diag}(600,450,600) ~kg\cdot m^2$. It is worth mentioning that the value of $\bm{J}_{\mathrm{c}0}$ is simply selected based on the known inertia matrix of the servicer. 

% \RV{In this paper, we initialize the log-value of hyperparameters based on the training data set $\mathcal{D}_N$: $\mathrm{log}\bm{\theta}_j^0=\mathrm{vec}(\mathrm{log}(\mathrm{std}(\bm{Y}_j/10)), ~\mathrm{log}(\mathrm{std}(\bm{Y}_j)),~\mathrm{log}(\mathrm{std}(\tilde{\bm{X}})))$. In \cite{chen2018priors}, it \TR{has been} illustrated that although the initial guess for the hyperparameters may have influences on the optimizing results, its impact on the regression accuracy of the GP model is almost negligible. Therefore, it is generally advisable to choose a relatively simple initial guess. }

\YLnew{For the obtained data set $\mathcal{D}_N${,} both a sparse and a standard GP are trained, of which the hyperparameters of both are optimized by the conjugated gradient descent algorithm. Specifically, in the case of ROSGP, the inducing points are initialized randomly inside $\mathcal{D}_N$ with the size of $M=50$ and the initial $\bm{\alpha}[0]$ is computed according to {\eqref{31}}. %its definition. 
%At the same time, setting 
{The} initial value of $\bm{P}$ {is set to} $\bm{P}[0]=10^2\bm{I}_M$. Then, at $t=50$s, the controller is updated to \eqref{ctrl} where the recursive update routine is activated.
The feedback gain functions $\bm{\zeta}_\mathrm{p}(\breve{k}_\mathrm{p},\bm{\Sigma}_\Delta)$  and $\bm{\zeta}_\mathrm{d}(\breve{k}_\mathrm{d},\bm{\Sigma}_\Delta)$  are selected as linear functions of the variance: $\bm{\zeta}_\mathrm{p}(\breve{k}_\mathrm{p},\bm{\Sigma}_\Delta) = \bm{J}_{\mathrm{c}0}(\breve{k}_\mathrm{p}+0.1{\bm{\Sigma}_\Delta^{1/2}})$ and $\bm{\zeta}_\mathrm{d}(\breve{k}_\mathrm{d},\bm{\Sigma}_\Delta) = \bm{J}_{\mathrm{c}0}(\breve{k}_\mathrm{d}+0.2{\bm{\Sigma}_\Delta^{1/2}})$ with $\breve{k}_\mathrm{p}=0.02$ and $\breve{k}_\mathrm{d}=0.05$. 
%\LY{These selections are { based on prior information, such as $\bm{J}_{\mathrm{c}0}$}. 
It can be easily verified that $\bm{\zeta}_\mathrm{p}(\cdot)$ and $\bm{\zeta}_\mathrm{d}(\cdot)$ satisfy Condition \ref{cond:bound} because the GP predictive variance is bounded by $\sup k_{**,j}$ for $j\in\mathbb{I}_1^3$.
Furthermore, the selection of parameters $\breve{k}_\mathrm{p}$ and $\breve{k}_\mathrm{d}$ is based on practical experience with \YLRV{PD control design for such systems with known nominal inertia matrix $\bm{J}_{\mathrm{c0}}$}, and does not require extensive tuning procedures. }
%It is only necessary to ensure that they constitute an admissible controller, i.e., a controller capable of stabilizing the system.

%It is worth mentioning that, the values of $\breve{k}_\mathrm{p}$ and $\breve{k}_\mathrm{d}$ are selected with practical experiences based on the prior knowledge of the combined spacecraft dynamics.}

{The} proposed recursive online sparse GP-based learning controller is compared to:
\begin{itemize} 
\item[1)] 
\YLnew{Baseline PD controller: This is the initial controller to generate the training data set during the first 50s, and it is kept fixed after $t=50$s.}

% Comparing the baseline PD controller with the proposed controller can clearly show the superior performance of the proposed method in terms of the steady-state error.
\item[2)] 
\YLnew{Standard GP-based controller: This controller keeps the same structure and parameters as the proposed controller \eqref{ctrl}, but {it is based only on} an initial trained GP without the recursive online update strategy.} % is employed.}
\end{itemize} 

\begin{figure}[t]
\centering
\begin{minipage}{0.5\textwidth}
\centering
\includegraphics[width=\textwidth]{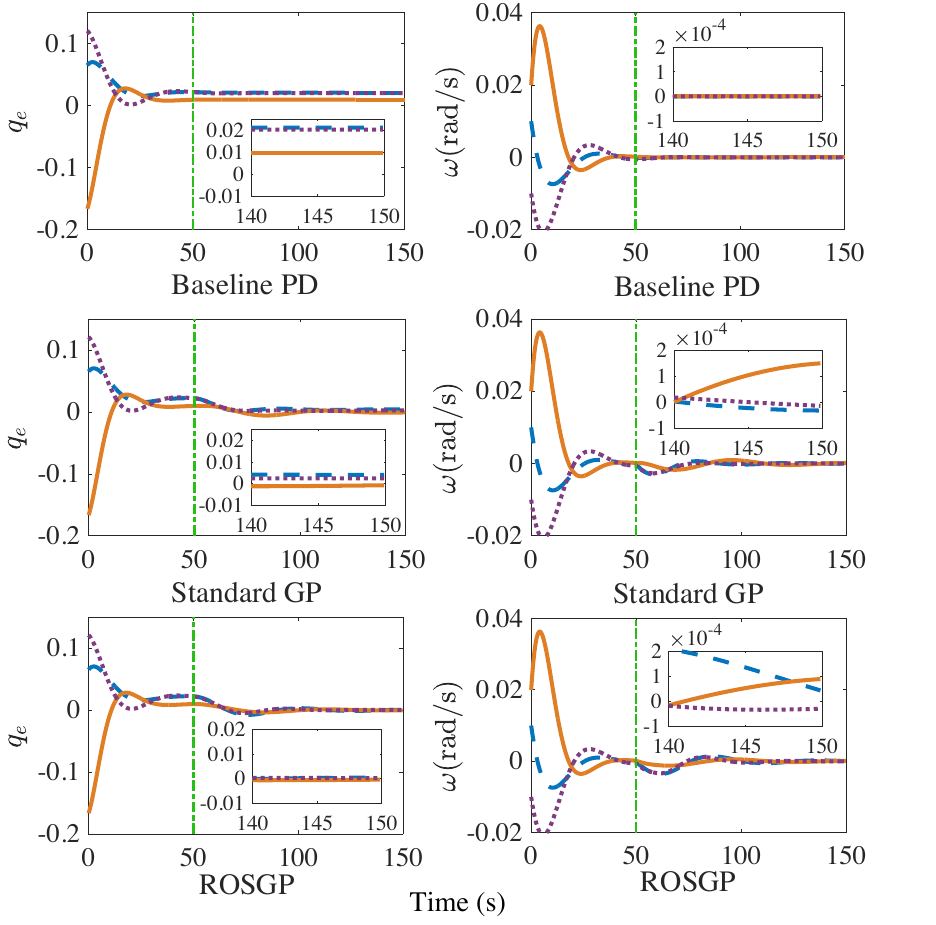}  
\caption[fig4]{\YLnew{Trajectories of $\bm{q}_\mathrm{e}$ and $\bm{\omega}$ under the considered controllers.} {The lines in each subplot describe the three components of these signals.} \YLRV{It can be observed that the ROSGP performs the best in terms of the steady-error of $\bm{q}_\mathrm{e}$ due to its capability of online GP compensation, while the baseline PD controller achieves the smallest steady-state error in terms of $\bm{\omega}$ at the expense of high-gain feedback.\vspace{-4mm}} 
}
\label{fig:4}
\end{minipage}
\end{figure}

\begin{figure}[t]
\centering
\begin{minipage}{0.5\textwidth}
\centering
\includegraphics[width=\textwidth]{figs/fig_4_5.pdf}  
\caption[fig4-5]{\RV{Control inputs $\bm{u}$ and the norm of feedback gain under the considered controllers. {It can be observed that the
high-gain feedback of the baseline PD controller remains
fixed throughout the entire task, while the GP-based controllers can adaptively adjust the
feedback gain according to the GP predictive variance,
keeping it at a low level while ensuring pointing accuracy.}\vspace{-4mm}} %however, it is at the expense of 
%high-gain feedback.}
}
\label{fig:4-5}
\end{minipage}
\end{figure}

\begin{figure}[t]
\centering
\begin{minipage}{0.45\textwidth}
\centering
\includegraphics[width=\textwidth]{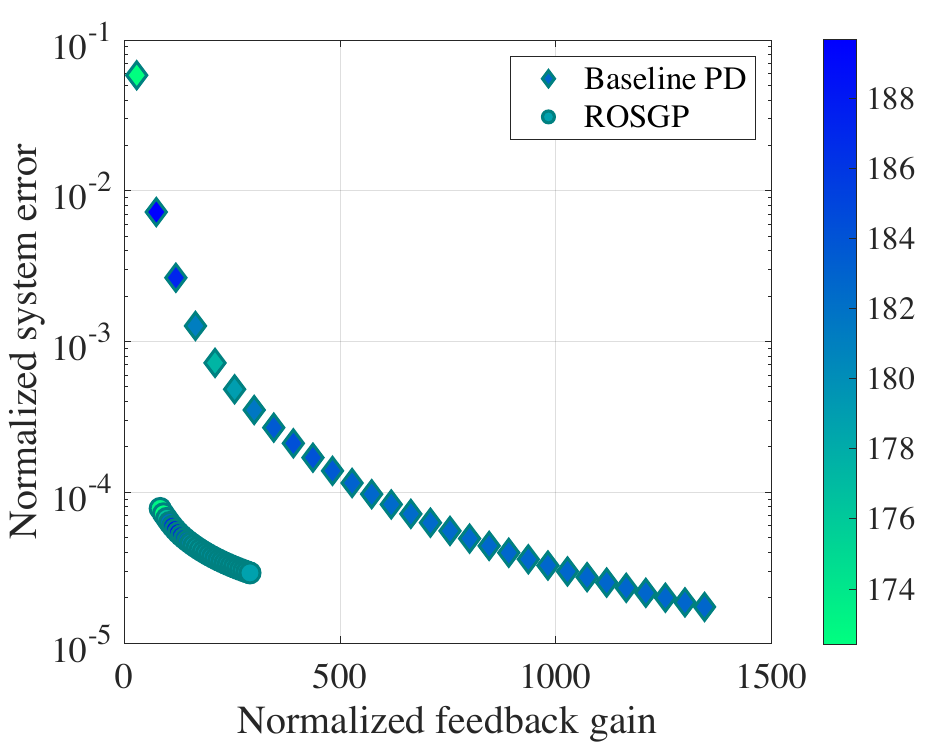}  
\caption[fig4-8]{\RV{A Pareto-front-like comparison of the normalized feedback gain and system error between the baseline PD and proposed ROSGP-based controllers. The color indicates the \YLRV{control efforts during the task.} It is apparent that at the same level of system error, the ROSGP-based controller requires a relatively small feedback gain.
}}
%however, it is at the expense of 
%high-gain feedback.}
\label{fig:4-8}
\end{minipage}
\end{figure}

\begin{figure}[t]
\centering
\begin{minipage}{0.45\textwidth}
\centering
\includegraphics[width=\textwidth]{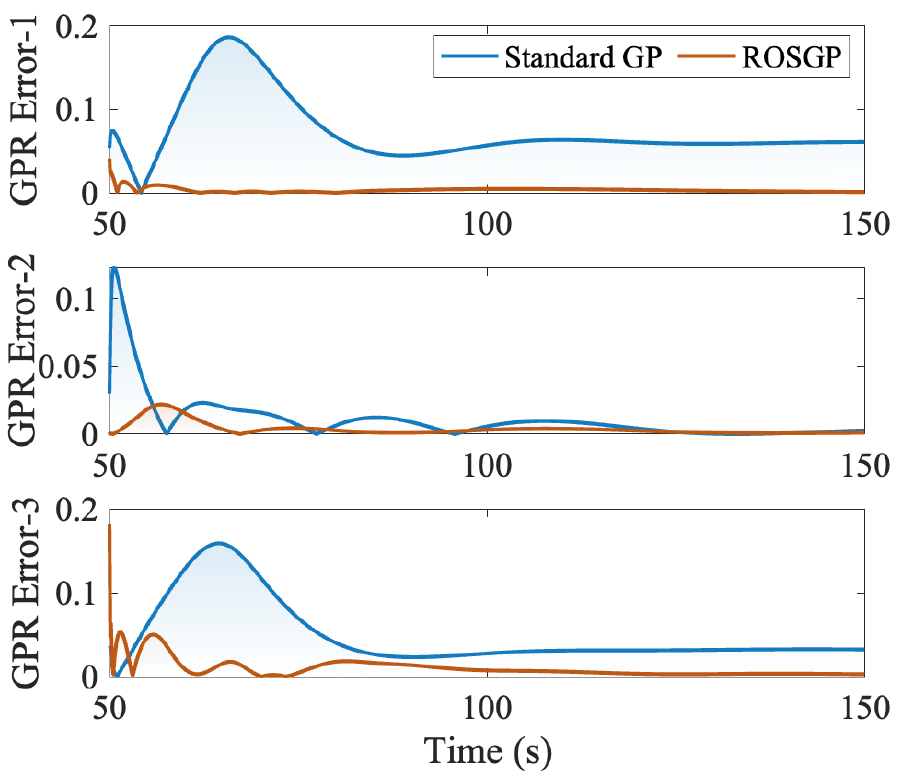}  
\caption[fig4]{\YLnew{Absolute value of the GP estimation error $\|\mu_{\Delta,j}-\breve{\Delta}_{j}\|$ for each dimension $j$ evaluated over time under the stabilization scenario.}}\vspace{-4mm}
\label{fig:5}
\end{minipage}
\end{figure}

\RV{The trajectories of the system states, control inputs, together with the feedback gain are depicted in Figs.~\ref{fig:4}-\ref{fig:4-5}, where the green line indicates that the controller updates at 50s. Simulation results show that all three controllers succeed in achieving the attitude stabilization task for the combined spacecraft. \YLRV{Also,  as shown in Fig.~\ref{fig:4-5}, the feedback gain of baseline PD remains constant during the entire task, and the two GP-based controllers can adapt its gain with the predictive variance. It is worth mentioning that the ROSGP performs the best in terms of final pointing error, because the unknown function $\breve{\bm{\Delta}}(\tilde{\bm{x}})$ in the combined spacecraft dynamics is real-time compensated by the predictive mean $\bm{\mu}_{\Delta}(\tilde{\bm{x}})$ of the trained GP model and further completed by the ROSGP algorithm with online adaption. 
However, it is more cautious hence its settling time is slightly slower which can be seen in terms of the slower convergence of the angular velocity. Additionally, while the baseline PD controller remains the largest steady-state error of $\bm{q}_\mathrm{e}$ without the GP compensation, it seems to achieve the smallest steady-state error of $\bm{\omega}$, nonetheless, at the expense of high-gain feedback. }
This can also be seen in Fig. ~\ref{fig:4-8}, where the Pareto-front-like figure shows the relationship between \YLRV{the normalized feedback gain $\sqrt{\|\bm{\zeta}_\mathrm{p}(\cdot)\|^2+\|\bm{\zeta}_\mathrm{d}(\cdot)\|^2}$ and system error $\sqrt{\|\bm{q}_\mathrm{e}\|^2+\|\bm{\omega}\|^2}$ of the two controllers in 30 simulation cases. The color indicates the control efforts $\int_{50}^{150}\|\bm{u}\|\mathrm{d}t$ during the task. It is clear that the proposed ROSGP-based controller requires a relatively modest feedback gain at the same level of system error.} }

%Moreover, compared with the baseline PD controller and the standard GP-based controller, the proposed ROSGP-based controller ensures the quaternion error $\bm{q}_\mathrm{e}$ finally converges into a smaller set while suffering from the presence of the unknown model and attitude maneuverability of the target. This is because the unknown function $\breve{\bm{\Delta}}(\tilde{\bm{x}})$ in the combined spacecraft dynamics is real-time compensated by the predictive mean $\bm{\mu}_{\Delta}(\tilde{\bm{x}})$ of the trained GP model and further completed by the ROSGP algorithm with online adaption. 
% Moreover, one can observe from Fig.~\ref{fig:4} that the proposed ROSGP-based controller further improves the dynamic response and steady-state error of the closed-loop system. The reason is that the unlearnt part in the standard GP model is further completed by the ROSGP algorithm with online adaption. 

The absolute value of the two GP estimation errors $\|\mu_{\Delta, j}-\breve{\Delta}_j\|$ is depicted in Fig.~\ref{fig:5}. One can see that there exists a significant estimation error between the predictive mean of standard GP and true function due to the initial data set $\mathcal{D}_N$ collected in the first 50s is not sufficient to describe the whole state space. In contrast, the model accuracy is adaptively enhanced by the proposed ROSGP, and the full compensation for the unknown function $\breve{\bm{\Delta}}(\tilde{\bm{x}})$ is achieved.

\subsection{Adaptive Control Under Re-maneuver Scenario}
During the on-orbit takeover control tasks, the combined spacecraft often needs to perform additional new tasks after attitude stabilization, such as maneuvering to a new attitude orientation for awaiting further missions. This new attitude orientation may be far from the initial training data set $\mathcal{D}_N$, which poses a significant challenge for the GP-based learning control strategy. Therefore, it is essential to verify the generalization performance and control effectiveness of the proposed GP-based online learning control strategy in untrained areas.

Consider the following scenario: On the basis of the effective attitude stabilization in the first 150s, the combined spacecraft is required to re-maneuver to a new attitude orientation $\bm{Q}_\mathrm{d}=[0.899~ -0.30~ 0.20~ -0.10]^{\top}$.  \RV{\YLRV{Meanwhile, to further illustrate the superior performance of the proposed ROSGP-based control strategy, rougher condition\TR{s} \TR{are} considered in this scenario.  In addition to larger uncertainties resulting from the re-orientation outside
the trained area and competitive torque of the target maneuver, large time-varying external disturbances that may result from the movement of an extra robotic arm, fuel sloshing, and flexible vibrations of solar panels, etc., are applied to the combined spacecraft from 150s,
%a negative effect that may result from the movement of an extra robotic arm, fuel sloshing, and flexible vibrations of solar panels, etc, is applied to the combined spacecraft from 150s to imitate the time-varying behaviors 
and can be modeled as}
$\bm{\tau}_\mathrm{d}(t)=[0.5\sin0.1t,-\sin0.15t,1.5\sin(-0.15t+1.5)]^{\top}\rm{N\cdot m}$.}
%\RV{Meanwhile, to further evaluate the performance of the proposed control strategy under time-varying external disturbances, a negative effect that may result from the movement of an extra robotic arm, fuel sloshing, and flexible vibrations of solar panels, etc, is applied to the combined spacecraft from 150s to imitate the time-varying behaviors and can be modeled as
%we assume that an extra robotic arm starts working on the target from 150s, where the negative effect can be modeled as 
% the combined spacecraft also starts to suffer from external sinusoidal disturbance torques in the form of 
%$\bm{\tau}_\mathrm{d}(t)=[0.5\sin0.1t,-\sin0.15t,1.5\sin(-0.15t+1.5)]^{\top}\rm{N\cdot m}$.}

\RV{
To show the advantages of the proposed GP-based learning control strategy, the NN-based adaptive control scheme (denoted by ANN) in \cite{huang2016impact} is applied in the considered scenario.  In this ANN approach, radial basis functions (RBF)  are chosen as the activation functions to estimate and compensate the unknown uncertainties. \YLRV{The NN structure $\hat{\bm{\Delta}}=\hat{\bm{\Theta}}^{\top}\bm{\Phi}_\rho$ used to estimate the unknown dynamics $\breve{\bm{\Delta}}$ 
contains 10 neurons with centers $c_i\in\mathbb{R}^6$ ($i \in \mathbb{I}_1^{10}$) randomly distributed in $[-2,2]^6$ and widths $\sigma_i=2$. 
Each element of the initial NN weights $\hat{\bm{\Theta}}(0)$ is chosen between $\pm 0.5$ randomly. The adaptive law for the NN weight is given by: $\dot{\hat{\Theta}}=\bm{F}_\rho \Phi_\rho \bm{r}^{\top}-k_\rho \bm{F}_\rho\|\bm{r}\| \hat{\Theta}$ where $\bm{r} = \bm{\omega}+\epsilon\bm{q}_\mathrm{e}$. The parameters for the ANN controller are selected as $\bm{F}_{\rho} = 200\bm{I}_3$, $k_\rho=1$, $\bm{\Lambda} = 0.1\bm{I}_3$, $\epsilon=0.1$.} \YLRV{It should be emphasized that the choice of all the aforementioned parameters in the ANN method is required to be tuned manually, for which there are no comprehensive tuning
rules. } For a fair comparison, the previously used control scenario was used where the GP-based feedforward has been substituted by the ANN compensation term.
%the same simulation is repeated by using the proposed control strategy where the ROSGP-based feedforward is substituted by the ANN compensation term.
}

\begin{figure}[t]
\centering
\begin{minipage}{0.5\textwidth}
\centering
\includegraphics[width=\textwidth]{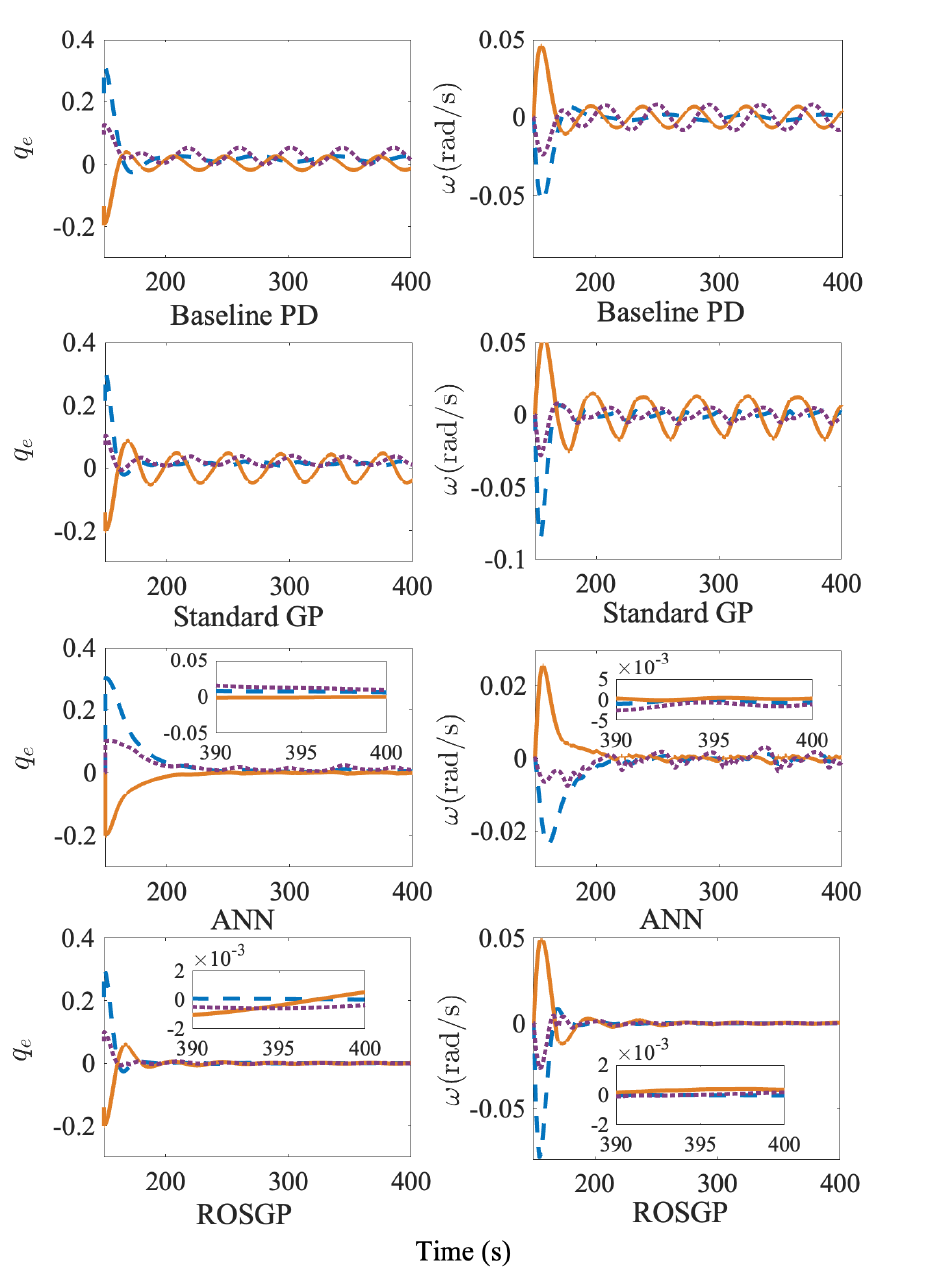}  
\caption[fig6]{\RV{Trajectories of $\bm{q}_\mathrm{e}$ and $\bm{\omega}$ under the considered controllers {and additional torque disturbances}. {Due to the torque disturbances, only ROSGP can achieve reasonably low steady-state error.}} \vspace{-4mm}} %\YLnew{The trajectories under the baseline PD controller seem to have a smaller steady-state error than the standard GP-based controller, however, it is at the expense of high-gain feedback.}}
\label{fig:6}
\end{minipage}
\end{figure}

\begin{figure}[t]
\centering
\begin{minipage}{0.5\textwidth}
\centering
\includegraphics[width=\textwidth]{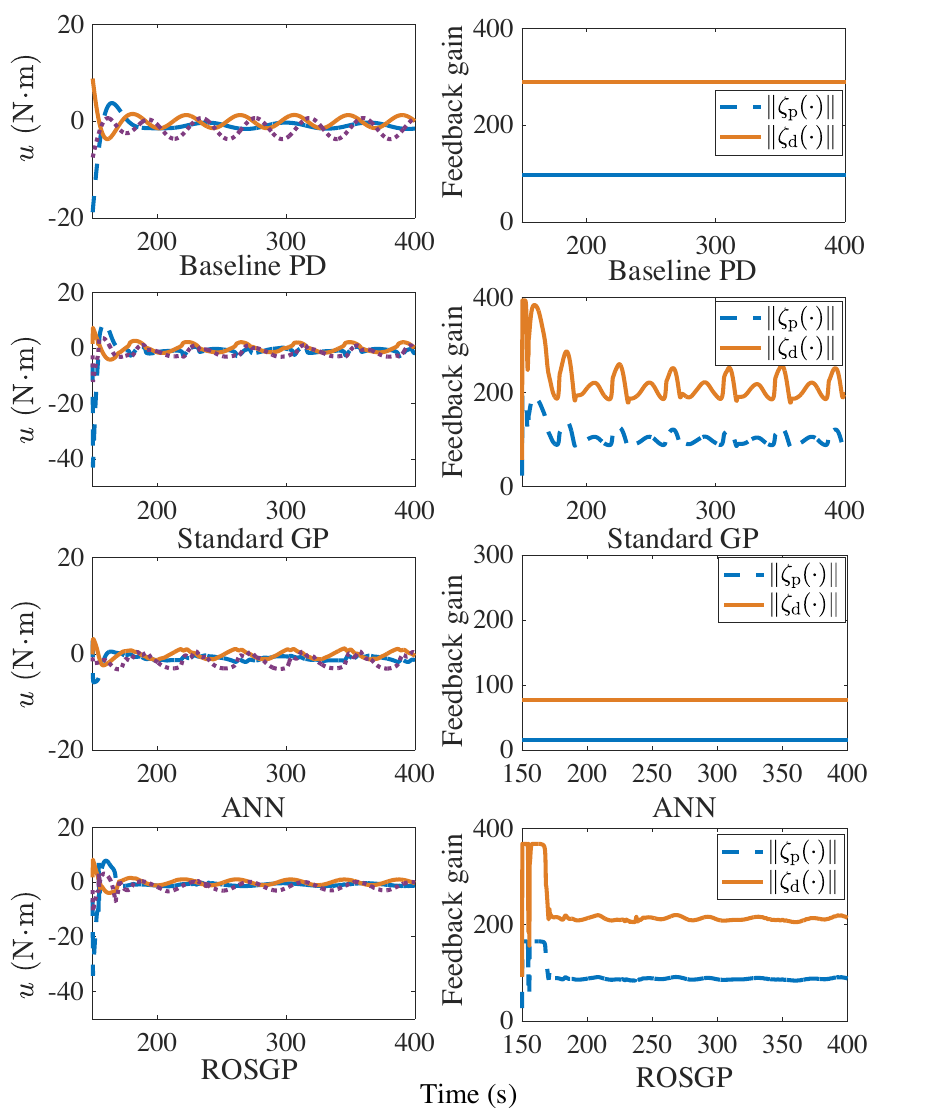}  
\caption[fig6-5]{\RV{Control inputs $\bm{u}$ and the norm of the feedback gain under the considered controllers. One can observe that the feedback gain of GP-based controllers can vary according to the predictive variance of GP in the untrained areas. \vspace{-4mm}} %however, it is at the expense of 
%high-gain feedback.}
}
\label{fig:6-5}
\end{minipage}
\end{figure}

\begin{figure}[htbp]
\centering
\begin{minipage}{0.49\textwidth}
\centering
\includegraphics[width=\textwidth]{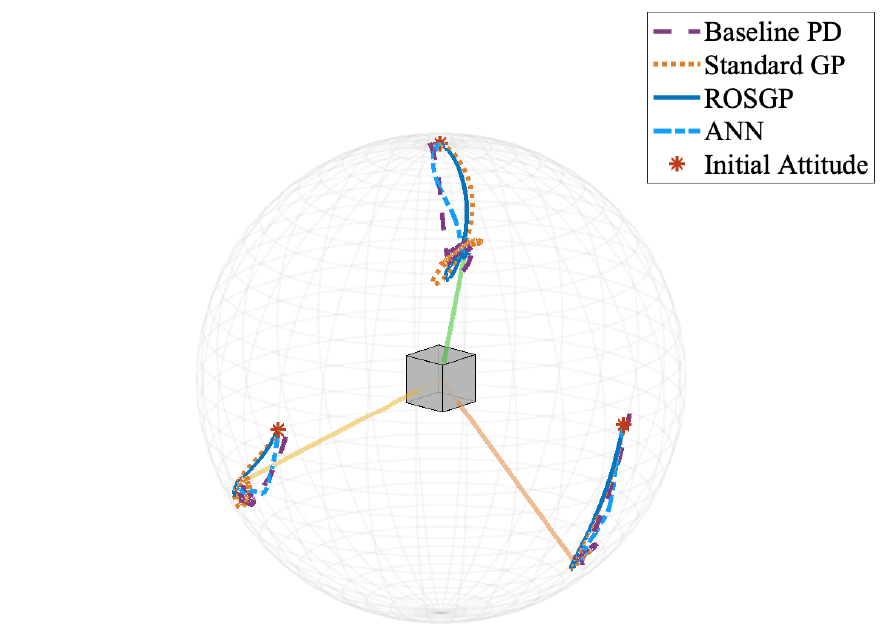}
\caption[fig7]{3-D trajectories of the spacecraft under the considered controllers. \vspace{-4mm}}
\label{fig:7}
\end{minipage}
\end{figure}

\begin{figure}[htbp]
\centering
\begin{minipage}{0.49\textwidth}
\centering
\includegraphics[width=\textwidth]{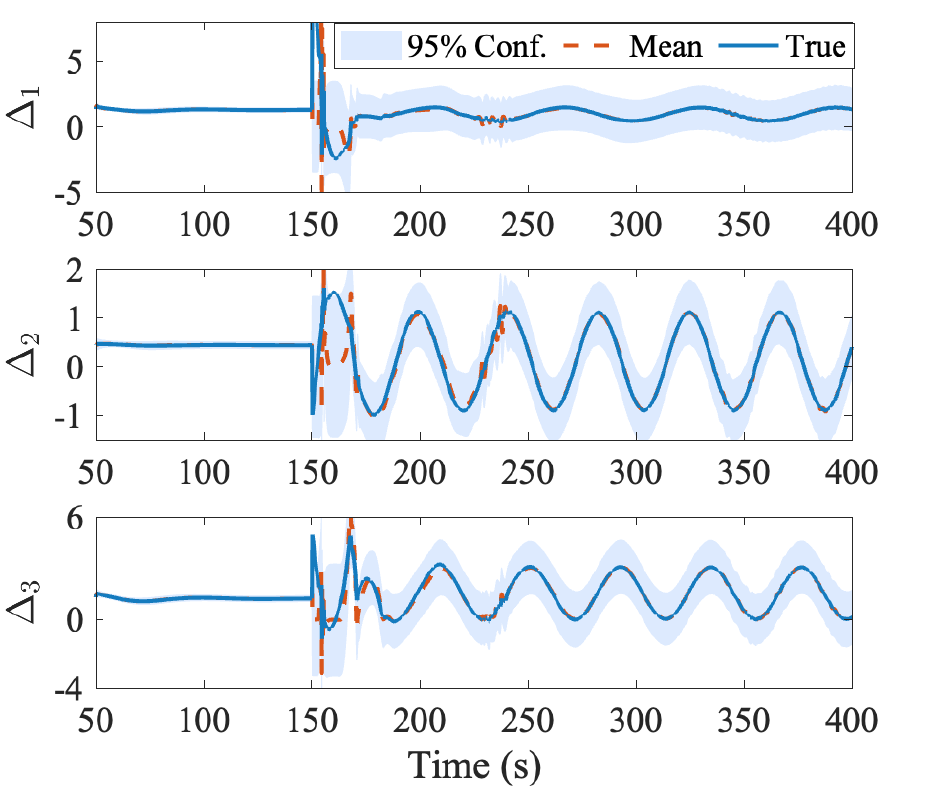}
\caption[fig9]{\RV{Value of the true uncertain dynamics (given by blue) {for each angular acceleration}, the mean of the learnt ROSGP  (given by red), and the 95\% confidence interval (given by shaded blue) evaluated at the true state values at time $t$ when re-maneuvers to the untrained area.}}
\label{fig:9}
\end{minipage}
\end{figure}

% \begin{table}[]
% \caption{{Feedback gain ranges for controllers.}}
% \centering
% \small
% \begin{tabular}{llllll}
% \hline
% \specialrule{0em}{1.0pt}{1.0pt}
% \hline
% \multicolumn{3}{c}{\multirow{2}{*}{Parameters}}             & \multicolumn{3}{c}{Value}                 \\ \cline{4-6} 
% \multicolumn{3}{c}{}                                & Baseline   & Standard GP        & ROSGP   \\ \hline
% \multirow{2}{*}{$\|\bm{\zeta}_\mathrm{p}(\cdot)\|$}      & \multicolumn{2}{c}{Range} & 96.05  & 21.73-192.54 & 24.17-112.68 \\ \cline{2-6} 
%                          & \multicolumn{2}{c}{Median}  & 96.05  &  84.93          & \textbf{64.74}        \\ \hline
% \multirow{2}{*}{$\|\bm{\zeta}_\mathrm{d}(\cdot)\|$}      & \multicolumn{2}{c}{Range} & 288.14 & 53.06-394.25   & 57.92-262.79  \\ \cline{2-6} 
%                          & \multicolumn{2}{c}{Median}  & 288.14 & 178.98         & \textbf{167.21}       \\ \hline
% \end{tabular}
% \label{tab:3}
% \end{table}

\RV{The trajectories of the system states, control inputs, together with the feedback gain under these settings are depicted in Figs.~\ref{fig:6}-\ref{fig:6-5}. \YLRV{It can be seen that under the additional disturbances,
%the four controllers can still achieve the goal of attitude re-maneuver under external time-varying disturbances. 
%Wherein, 
both the dynamic response and steady-state error of the baseline PD and standard GP-based schemes are obviously unsatisfactory and \TR{fail to sufficiently well} perform the attitude re-orientation task, \TR{resulting in} %which depicts 
an oscillating behavior around the equilibrium points. The reason is that the high-gain feedback of the baseline PD controller is not capable of dealing with such large uncertainties anymore, which also cannot be precisely captured by the offline-trained standard GP model. The performance of the proposed ROSGP-based controller is significantly \TR{better} %improved 
compared %with
\TR{to} the baselines, mainly resulting from the
%introduction of GP compensation together with the 
\TR{online adaptation of the GP compensation}  strategy. 
Particularly, compared with ANN method, the system states converge faster under the proposed ROSGP-based controller because of the adaptive feedback gains with GP predictive variance.}}
A 3D illustration of the trajectories of the axes of $\mathcal{F}_\mathrm{B}$ is shown in Fig.~\ref{fig:7}, where the gray cube in the center represents the combined spacecraft. \YLRV{One can observe that the trajectories under the baseline PD controller show an obvious limit cycle behavior around the desired orientation. 
This phenomenon can be further demonstrated by Fig.~\ref{fig:9}.}
%and Table.~\ref{tab:3}.}

\YLRV{As shown in Fig.~\ref{fig:9}, the predictive variance (depicted by a shaded area of 95\% confidence interval) keeps at a low level from 50s to 150s during the stabilization task. Next, the predictive variance increases significantly from 150s when the combined spacecraft re-maneuvers to an area outside the training data set. }
 \RV{This makes it so that
%renders
the feedback gain of the two GP-based learning controllers appropriately increases to \YLRV{further mitigate the suddenly appearing time-varying disturbance.} }
%Meanwhile,  Table~\ref{tab:3} {shows} that, compared with the high-gain feedback of the baseline controller that remains fixed throughout the entire task, the standard GP and the ROSGP-based controllers can adaptively adjust the feedback gain according to the GP predictive variance, keeping it at a low level while ensuring {pointing} accuracy. Note that the upper bound of the feedback gains of the two GP-based controllers is relatively high. This is due to the {transient of the} controller {adaptation,} %and the fluctuation of  
%{corresponding to} the convergence of $\bm{\alpha}$ in the initial phase of the online update algorithm. Once $\bm{\alpha}$ {has converged}, the feedback gain returns to a lower level.

\begin{figure}[t]
\centering
\begin{minipage}{0.47\textwidth}
\centering
\includegraphics[width=\textwidth]{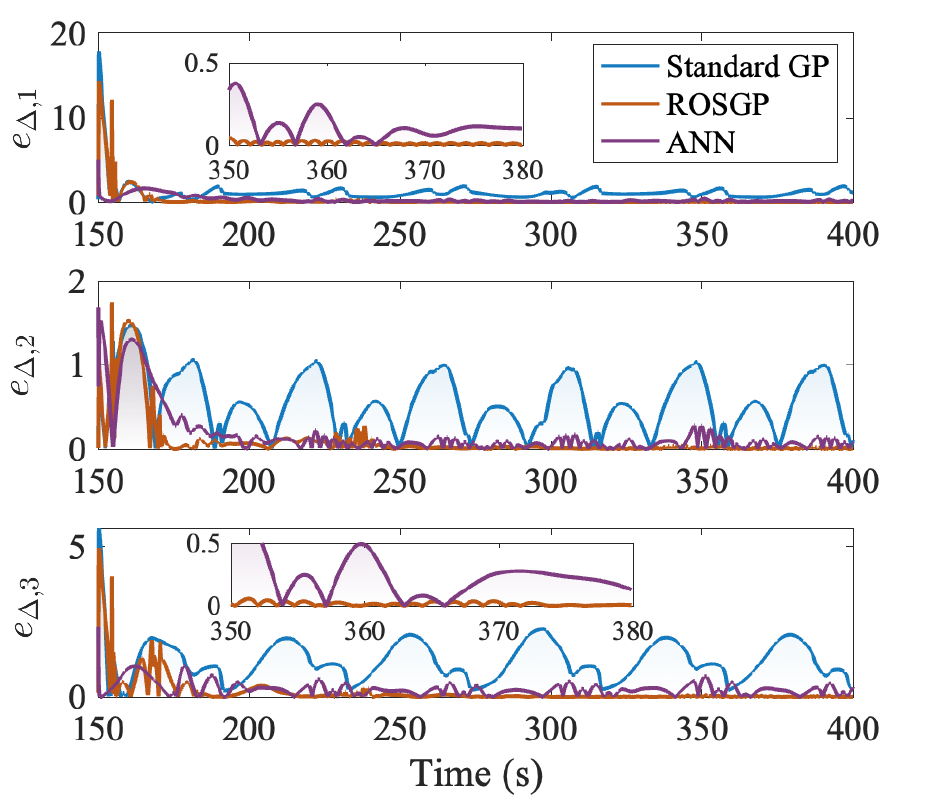}
\caption[fig8]{\RV{Absolute value of the  estimation error $e_{\Delta,j}$ for each dimension $j$ under the considered controllers evaluated over time under the re-orientation scenario.}}
\label{fig:8}
\end{minipage}
\end{figure}

% \begin{figure}[htbp]
% \centering
% \begin{minipage}{0.40\textwidth}
% \centering
% \includegraphics[width=\textwidth]{figs/fig_10.pdf} 
% \caption[fig10]{Computation time for {a} $400$s long simulation with the considered controllers.}
% \label{fig:10}
% \end{minipage}
% \end{figure}

\RV{
Figure \ref{fig:8} shows the absolute value of the estimation error $e_{\Delta,j}=\|\mu_{\Delta, j}-\breve{\Delta}_j\|$ of the two GPs and $e_{\Delta,j}=\|\hat{\Delta}_j-\breve{\Delta}_j\|$ of the ANN \YLRV{starting from 150s, i.e., the re-maneuver task begins}. It can be observed that despite the presence of target attitude maneuvers, external time-varying disturbances, and unlearnt {dynamics}, the ROSGP still has the capability of online learning of complicated time-varying uncertainties and compensating for them. Therefore, \YLRV{despite realizing an attitude re-orientation in a previously unseen area of the state space}, the set point error can always be maintained at a low level with superior transient performance. It should be noted that the ANN method results in a comparable estimation error w.r.t. the proposed ROSGP, \YLRV{but this comes at the cost of a significant manual tuning of the hyperparameters of the ANN scheme.}}

In addition, the computation time for a 400s long simulation (on a MacOS Monterey, 10-core M1 Pro, 16GB RAM) under different controllers is illustrated 
% in Fig.~\ref{fig:10}. 
as follows. %Particularly, compared with the standard GP, 
\YLRV{Particularly, the time costs of the standard GP and ROSGP-based controller during model training {is $5.59$s and $2.71$s while during task execution of a $400$s long trajectory.  The average computation time of the standard GP, ROSGP, and ANN-based controller per control cycle is $16.36$ms, $3.31$ms, and $3.99$ms (which are realizable under the sampling time $T_\mathrm{s}=100$ms), respectively.}}  %are 2.71s and 13.25s, respectively, showing 
{This shows} that the computational load of the proposed control strategy is quite small, \YLRV{and does not require
extensive parameter tuning procedures.}
%Finally, Fig.~\ref{fig:11} %.(a) and (b) respectively 
%presents the steady-state errors of $\bm{q}_\mathrm{e}$ and $\bm{\omega}$ under the three considered controllers during the stabilization and {the} re-maneuver phases. Note that although the steady-state error of the baseline controller is relatively small in some cases, this comes at the cost of high-gain feedback as discussed above.
Thus, the results are consistent with the expected performance of the control system design, indicating that the proposed strategy can efficiently, yet accurately learn the unknown function online, and achieve high-precision control of the attitude takeover task.

% \begin{figure}[t]
% \centering
% \begin{minipage}{0.49\textwidth}
% \centering
% \includegraphics[width=\textwidth]{figs/fig_11.pdf}
% \caption[fig11]{Steady {state} error %of states and 
% under {the considered} controllers: (a) {Steady-state-error} of $\bm{q}_\mathrm{e}$. (b) {Steady-state-error} of $\bm{\omega}$. }
% \label{fig:11}
% \end{minipage}
% \end{figure}

\subsection{Monte Carlo Simulation}
Finally, Monte Carlo simulation is conducted to comprehensively analyze the effectiveness and generalization of the proposed control strategy under different {physical parameter} values and {user chosen} controller parameters. Table.~\ref{tab:2} presents the range of {the} selected random parameters in the Monte Carlo simulation.

\begin{table}[h]
\centering
\caption[table2]{{Randomized} parameter {ranges} in {the} Monte Carlo {study}.}
\begin{tabular}{ll}
\hline
\specialrule{0em}{1.0pt}{1.0pt}
\hline
Randomized Parameters                                           & Range \\ \hline
$m_\mathrm{t}$, kg                                          & [50,100] \\
$\bm{J}_\mathrm{t}$, kg$\cdot $m$^2$ & $\rm{diag}[50\pm50,~50\pm50,~50\pm50]$ \\
$\bm{Q}_{0}$                                  & ${\mathbb{Q}}^3$\\
$\bm{\omega}_{0}$, rad/s                 & [-0.1,0.1]$\times$[-0.1,0.1]$\times$[-0.1,0.1] \\
Baseline $\bm{K}_\mathrm{p} $                                  &  $[0.08,~0.25]\bm{J}_{\mathrm{c}0}$ \\
Baseline $\bm{K}_\mathrm{d}$                                    & $[0.28,0.55]\bm{J}_{\mathrm{c}0}$  \\
$\bm{P}[0]$                                      & [1,~1000]$\bm{I}_M$  \\
 $\bm{K}_\mathrm{pt}$                                           &  $[0.01, 0.3]\bm{J}_\mathrm{t} $\\
$\bm{K}_\mathrm{dt}$                                          &$[0.02, 0.8]\bm{J}_\mathrm{t} $ \\
Amplitude of $\bm{\tau}_\mathrm{d}(t)$                                    & [-0.5,0.5]$\times$[-0.5,0.5]$\times$[-0.5,0.5]\\ \hline
\end{tabular}
\label{tab:2}
\end{table}

\begin{figure}[htbp]
\centering
\begin{minipage}{0.49\textwidth}
\centering
\includegraphics[width=\textwidth]{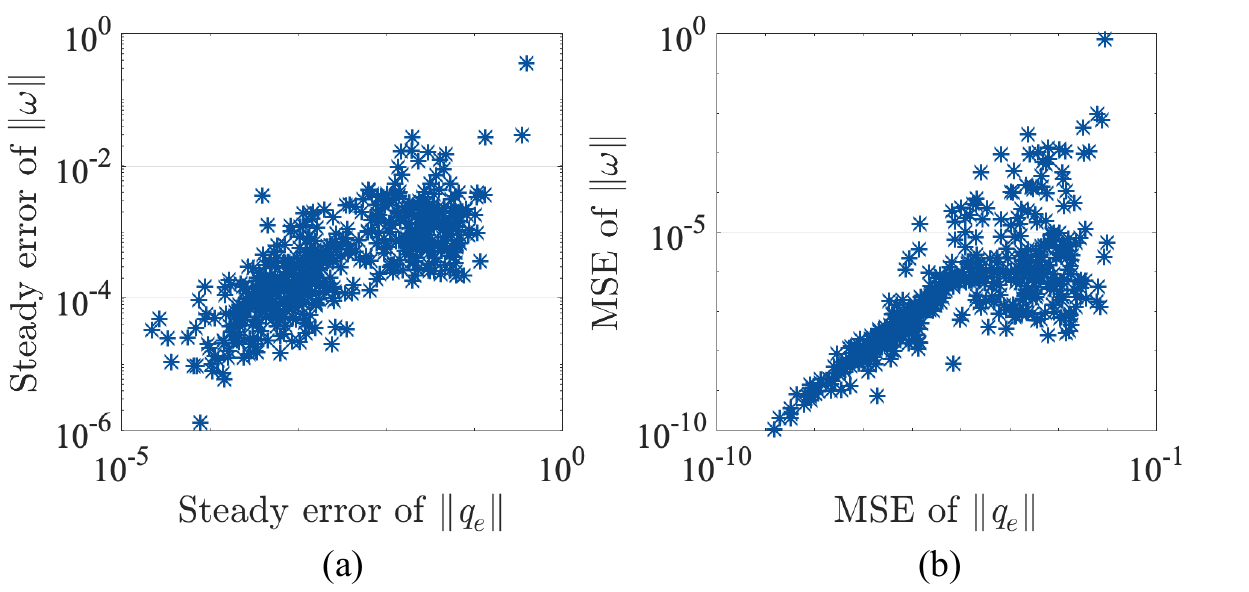}
\caption[fig12]{Distributions of (a) steady-state error and (b) MSE of $\bm{q}_\mathrm{e}$ and $\bm{\omega}$. }
\label{fig:12}
\end{minipage}
\end{figure}

Figure~\ref{fig:12} presents the distributions of the steady-state error and \emph{mean square error} (MSE) of $\bm{q}_\mathrm{e}$ and $\bm{\omega}$ from 500 Monte Carlo simulations, respectively, where each ``*" represents a simulation case under random conditions. It can be observed that the overwhelming majority of the simulation cases {correspond to good performance}, where the steady-state error of the attitude quaternion is less than $1\times 10^{-2}$ and that of attitude angular velocity is less than $1\times10^{-2}\rm{rad/s}$. Additionally, the MSE of attitude quaternion is less than $1\times10^{-4}$, and that of attitude angular velocity is less than $1\times10^{-4}\rm{rad/s}$. \RV{The results illustrate the generalization performance of the proposed ROSGP-based controller.}

%Further analyzing the simulation cases with relatively poor results shown in Fig.~\ref{fig:12}, one can find that these results are due to the exaggeratedly randomized conditions. For example, considering the highest point in Fig.~\ref{fig:12} (corresponding to the simulation case with the worst steady-state error and mean square error among the 500 Monte Carlo simulations), the parameters of the active attitude control law for the target are randomized as $\bm{K}_\mathrm{pt}=0.292 \bm{J}_\mathrm{t}, \bm{K}_\mathrm{dt}=0.796\bm{J}_\mathrm{t}$, while the initial matrix in ROSGP algorithm is randomized as $\bm{P}[0]=1.05\bm{I}_M$. This indicates a significant deviation between the unknown uncertainties in the current model and the nominal model, while the prior distribution defined by the initial parameter value $\bm{P}[0]$ is relatively small. This discrepancy is the reason why the steady-state error can not converge into a smaller set. 

In summary, the simulation results show that the proposed GP-based online learning control strategy presented in \eqref{ctrl} and Algorithm~\ref{alg:1} \YLRV{is highly effective and its implementation is feasible for OOS attitude takeover tasks} even in the presence of target attitude maneuverability.

\section{CONCLUSION}

This paper present{s} an effective GP-based online learning control strategy for attitude takeover control of noncooperative targets with attitude maneuverability. A novel recursive online sparse GP algorithm {is introduced} to ensure the successive online learning of the unknown dynamics, while the computational load {is} kept low, making {the approach well applicable} %efficient for use 
in resource-constrained onboard scenarios. The proposed method has probabilistic guarantees for a user-defined bound of the {pointing error}. The introduced approach provides new perspectives into the attitude controller design of spacecraft{s} with unknown dynamics, especially in cases where the moment of inertia cannot be identified. The properties and effectiveness of our proposed strategy have been analyzed and demonstrated by numerical simulations based on a high-fidelity simulator.

\bibliographystyle{IEEEtaes}
\bibliography{reference}

\begin{thebibliography}{10}
\providecommand{\url}[1]{#1}
\csname url@samestyle\endcsname
\renewcommand{\newblock}{\par}
\providecommand{\bibinfo}[2]{#2}
\providecommand{\BIBentrySTDinterwordspacing}{\spaceskip=0pt\relax}
\providecommand{\BIBentryALTinterwordstretchfactor}{4}
\providecommand{\BIBentryALTinterwordspacing}{\spaceskip=\fontdimen2\font plus
\BIBentryALTinterwordstretchfactor\fontdimen3\font minus
  \fontdimen4\font\relax}
\providecommand{\BIBforeignlanguage}[2]{{%
\expandafter\ifx\csname l@#1\endcsname\relax
\typeout{** WARNING: IEEEtran.bst: No hyphenation pattern has been}%
\typeout{** loaded for the language `#1'. Using the pattern for}%
\typeout{** the default language instead.}%
\else
\language=\csname l@#1\endcsname
\fi
#2}}
\providecommand{\BIBdecl}{\relax}
\BIBdecl

\bibitem{flores2014review}
A.~Flores-Abad, O.~Ma, K.~Pham, and S.~Ulrich
\newblock  A review of space robotics technologies for on-orbit servicing
  \newblock  \emph{Prog. Aerosp. Sci.}, vol.~68, pp. 1--26, 2014.

\bibitem{pinard2007accurate}
D.~Pinard, S.~Reynaud, P.~Delpy, and S.~E. Strandmoe
\newblock  Accurate and autonomous navigation for the {ATV} \newblock
  \emph{Aerosp. Sci. Technol.}, vol.~11, no.~6, pp. 490--498, 2007.

\bibitem{liu2018key}
H.~Liu, Z.~Li, Y.~Liu, M.~Jin, F.~Ni, and Y.~Liu
\newblock  Key technologies of {T}ian{G}ong-2 robotic hand and its on-orbit
  experiments \newblock  \emph{Sci. Sin. Technol.}, vol.~48, no.~12, pp.
  1313--1320, 2018.

\bibitem{bergmann1987mass}
E.~Bergmann, B.~K. Walker, and D.~R. Levy
\newblock  Mass property estimation for control of asymmetrical satellites
  \newblock  \emph{J. Guid., Control, Dyn.}, vol.~10, no.~5, pp. 483--491,
  1987.

\bibitem{murotsu1994parameter}
Y.~Murotsu, K.~Senda, M.~Ozaki, and S.~Tsujio
\newblock  Parameter identification of unknown object handled by free-flying
  space robot \newblock  \emph{J. Guid., Control, Dyn.}, vol.~17, no.~3, pp.
  488--494, 1994.

\bibitem{ma2008orbit}
O.~Ma, H.~Dang, and K.~Pham
\newblock  On-orbit identification of inertia properties of spacecraft using a
  robotic arm \newblock  \emph{J. Guid., Control, Dyn.}, vol.~31, no.~6, pp.
  1761--1771, 2008.

\bibitem{christidi2023parameter}
O.-O. Christidi-Loumpasefski and E.~Papadopoulos
\newblock  On the parameter identification of free-flying space manipulator
  systems \newblock  \emph{Robot. Auton. Syst.}, vol. 160, p. 104310, 2023.

\bibitem{christidi2023system}
O.-O. Christidi-Loumpasefski, G.~Rekleitis, E.~Papadopoulos, and F.~Ankersen
\newblock  On system identification of space manipulator systems including
  their fuel sloshing effects \newblock  \emph{IEEE Robot. Autom. Lett.},
  vol.~8, no.~5, pp. 2446--2453, 2023.

\bibitem{meng2019identification}
Q.~Meng, J.~Liang, and O.~Ma
\newblock  Identification of all the inertial parameters of a non-cooperative
  object in orbit \newblock  \emph{Aerosp. Sci. Technol.}, vol.~91, pp.
  571--582, 2019.

\bibitem{chu2020least}
W.~Chu, S.~Wu, Z.~Wu, and Y.~Wang
\newblock  Least square based ensemble deep learning for inertia tensor
  identification of combined spacecraft \newblock  \emph{Aerosp. Sci.
  Technol.}, vol. 106, p. 106189, 2020.

\bibitem{huang2015adaptive}
P.~Huang, D.~Wang, Z.~Meng, F.~Zhang, and J.~Guo
\newblock  Adaptive postcapture backstepping control for tumbling tethered
  space robot-target combination \newblock  \emph{J. Guid., Control, Dyn.},
  vol.~39, no.~1, pp. 150--156, 2015.

\bibitem{zhang2017coordinated}
B.~Zhang, B.~Liang, Z.~Wang, Y.~Mi, Y.~Zhang, and Z.~Chen
\newblock  Coordinated stabilization for space robot after capturing a
  noncooperative target with large inertia \newblock  \emph{Acta Astronaut},
  vol. 134, pp. 75--84, 2017.

\bibitem{guohua2020adaptive}
G.~Kang, J.~Wu, C.~Jin, and X.~Chen
\newblock  Adaptive controller design for satellite attached by non-cooperative
  object \newblock  \emph{Chin. J. Aeronaut.}, vol.~33, no.~3, pp. 1006--1015,
  2020.

\bibitem{liu2021nonfragile}
C.~Liu, X.~Yue, and Z.~Yang
\newblock  Are nonfragile controllers always better than fragile controllers in
  attitude control performance of post-capture flexible spacecraft?
  \emph{Aerosp. Sci. Technol.}, vol. 118, p. 107053, 2021.

\bibitem{LEEGHIM2009778}
H.~Leeghim, Y.~Choi, and H.~Bang
\newblock  Adaptive attitude control of spacecraft using neural networks
  \newblock  \emph{Acta Astronaut}, vol.~64, no.~7, pp. 778--786, 2009.

\bibitem{huang2016impact}
P.~Huang, D.~Wang, Z.~Meng, F.~Zhang, and Z.~Liu
\newblock  Impact dynamic modeling and adaptive target capturing control for
  tethered space robots with uncertainties \newblock  \emph{IEEE/ASME Trans.
  Mechatronics}, vol.~21, no.~5, pp. 2260--2271, 2016.

\bibitem{ning2021event}
K.~Ning, B.~Wu, and C.~Xu
\newblock  Event-triggered adaptive fuzzy attitude takeover control of
  spacecraft \newblock  \emph{Adv. Space. Res}, vol.~67, no.~6, pp. 1761--1772,
  2021.

\bibitem{wei2018adaptive}
C.~Wei, J.~Luo, H.~Dai, and J.~Yuan
\newblock  Adaptive model-free constrained control of postcapture flexible
  spacecraft: a {E}uler-{L}agrange approach \newblock  \emph{J. Vib. Contr.},
  vol.~24, no.~20, pp. 4885--4903, 2018.

\bibitem{luo2018low}
J.~Luo, Z.~Yin, C.~Wei, and J.~Yuan
\newblock  Low-complexity prescribed performance control for spacecraft
  attitude stabilization and tracking \newblock  \emph{Aerosp. Sci. Technol.},
  vol.~74, pp. 173--183, 2018.

\bibitem{fan2021inertia}
Y.~Fan and W.~Jing
\newblock  Inertia-free appointed-time prescribed performance tracking control
  for space manipulator \newblock  \emph{Aerosp. Sci. Technol.}, p. 106896,
  2021.

\bibitem{rasmussen2003gaussian}
C.~E. Rasmussen
\newblock  Gaussian processes in machine learning \newblock  In \emph{Advanced
  Lectures on Machine Learning}. Berlin, Germany: Springer, 2004, pp. 63--71.

\bibitem{beckers2019stable}
T.~Beckers, D.~Kuli{\'c}, and S.~Hirche
\newblock  Stable {G}aussian process based tracking control of
  {E}uler--{L}agrange systems \newblock  \emph{Automatica}, vol. 103, pp.
  390--397, 2019.

\bibitem{liu2021learning}
Y.~Liu and R.~T{\'o}th
\newblock  Learning based model predictive control for quadcopters with dual
  {G}aussian process \newblock  In \emph{Proc. IEEE Conf. Decis. Control
  (CDC)}. Austin, TX, USA, December 14-17, 2021, pp. 1515--1521.

\bibitem{kabzan2019learning}
J.~Kabzan, L.~Hewing, A.~Liniger, and M.~N. Zeilinger
\newblock  Learning-based model predictive control for autonomous racing
  \newblock  \emph{IEEE Robot. Automat. Lett.}, vol.~4, no.~4, pp. 3363--3370,
  2019.

\bibitem{csato2002sparse}
L.~Csat{\'o} and M.~Opper
\newblock  Sparse on-line {G}aussian processes \newblock  \emph{Neural
  Comput.}, vol.~14, no.~3, pp. 641--668, 2002.

\bibitem{grande2014experimental}
R.~C. Grande, G.~Chowdhary, and J.~P. How
\newblock  Experimental validation of bayesian nonparametric adaptive control
  using {G}aussian processes \newblock  \emph{J. Aerosp. Inform. Syst.},
  vol.~11, no.~9, pp. 565--578, 2014.

\bibitem{chowdhary2014bayesian}
G.~Chowdhary, H.~A. Kingravi, J.~P. How, and P.~A. Vela
\newblock  Bayesian nonparametric adaptive control using {G}aussian processes
  \newblock  \emph{IEEE Trans. Neural Netw. Learn. Syst.}, vol.~26, no.~3, pp.
  537--550, 2014.

\bibitem{ignatyev2023sparse}
D.~I. Ignatyev, H.-S. Shin, and A.~Tsourdos
\newblock  Sparse online {G}aussian process adaptation for incremental
  backstepping flight control \newblock  \emph{Aerosp. Sci. Technol.}, vol.
  136, p. 108157, 2023.

\bibitem{petelin2011control}
D.~Petelin and J.~Kocijan
\newblock  Control system with evolving {G}aussian process models \newblock  In
  \emph{Proc. IEEE Workshop on Evolving and Adaptive Intelligent Systems
  (EAIS)}, 2011, pp. 178--184.

\bibitem{kocijan2016modelling}
J.~Kocijan
\newblock \emph{Modelling and control of dynamic systems using Gaussian process
  models}. Springer, 2016.

\bibitem{maiworm2021online}
M.~Maiworm, D.~Limon, and R.~Findeisen
\newblock  Online learning-based model predictive control with gaussian process
  models and stability guarantees \newblock  \emph{Int. J. Robust Nonlinear
  Control}, vol.~31, no.~18, pp. 8785--8812, 2021.

\bibitem{ma2021learning}
G.~Ma, Y.~Liu, Y.~Lyu, and P.~Wang
\newblock  Learning-based attitude takeover control for noncooperative space
  targets with unknown dynamics \newblock  In \emph{Proc. IEEE Chin. Control
  Conf. (CCC)}. Shanghai, China, July, 2021, pp. 2233--2238.

\bibitem{wie2008space}
B.~Wie
\newblock \emph{Space Vehicle Dynamics and Control}. Reston, VA, USA: AIAA
  Education Series, 2008.

\bibitem{chen2018priors}
Z.~Chen and B.~Wang
\newblock  How priors of initial hyperparameters affect {G}aussian process
  regression models \newblock  \emph{Neurocomputing}, vol. 275, pp. 1702--1710,
  2018.

\bibitem{burden2015numerical}
R.~L. Burden, J.~D. Faires, and A.~M. Burden
\newblock \emph{Numerical analysis}. 10th ed. Boston, MA, USA: Cengage
  Learning, 2014.

\bibitem{snelson2005sparse}
E.~Snelson and Z.~Ghahramani
\newblock  Sparse {G}aussian processes using pseudo-inputs \newblock
  \emph{Adv. Neural Inf. Process. Syst.}, vol.~18, pp. 1257--1264, 2005.

\bibitem{umlauft2018uncertainty}
J.~Umlauft, L.~P{\"o}hler, and S.~Hirche
\newblock  An uncertainty-based control {L}yapunov approach for control-affine
  systems modeled by {G}aussian process \newblock  \emph{IEEE Control Syst.
  Lett.}, vol.~2, no.~3, pp. 483--488, 2018.

\bibitem{srinivas2012information}
N.~Srinivas, A.~Krause, S.~M. Kakade, and M.~W. Seeger
\newblock  Information-theoretic regret bounds for {G}aussian process
  optimization in the bandit setting \newblock  \emph{IEEE Trans. Inf. Theory},
  vol.~58, no.~5, pp. 3250--3265, 2012.

\bibitem{umlauft2019feedback}
J.~Umlauft and S.~Hirche
\newblock  Feedback linearization based on {G}aussian processes with
  event-triggered online learning \newblock  \emph{IEEE Trans. Automat.
  Control}, vol.~65, no.~10, pp. 4154--4169, 2019.

\bibitem{gui2018robustness}
H.~Gui and A.~H. de~Ruiter
\newblock  Robustness analysis and performance tuning for the quaternion
  proportional--derivative attitude controller \newblock  \emph{J. Guid.,
  Control, Dyn.}, vol.~41, no.~10, pp. 2308--2317, 2018.

\bibitem{huang2016reconfigurable}
P.~Huang, M.~Wang, Z.~Meng, F.~Zhang, Z.~Liu, and H.~Chang
\newblock  Reconfigurable spacecraft attitude takeover control in post-capture
  of target by space manipulators \newblock  \emph{J. Franklin Inst.}, vol.
  353, no.~9, pp. 1985--2008, 2016.

\end{thebibliography}

\end{document}